\documentclass[aps,pra,preprint,superscriptaddress,nofootinbib]{revtex4-1} 

\usepackage{amsmath,amssymb,amsfonts}
\usepackage{psfrag}
\usepackage{color}
\definecolor{darkblue}{rgb}{0.1,0.1,.7}
\usepackage[colorlinks, linkcolor=darkblue, citecolor=darkblue, urlcolor=darkblue, linktocpage]{hyperref} 
\usepackage{natbib}

\usepackage[]{amsmath}
\usepackage[]{graphicx}
\usepackage[]{latexsym}
\usepackage{geometry}
\usepackage{amscd}
\usepackage[all,cmtip]{xy}
\usepackage{mathrsfs}
\usepackage{simplewick}
\usepackage{changepage}
\numberwithin{equation}{section}

\newcommand{\bea}{\begin{eqnarray}}
\newcommand{\eea}{\end{eqnarray}}

\def\beq{\begin{equation}} 
\def\eeq{\end{equation}} 
\def\<{\langle}
\def\>{\rangle}

\def\cO {{\cal O}}

\newcommand{\vx}{\vec{x}}
\newcommand{\ket}[1]{\left| #1 \right>} 
\newcommand{\bra}[1]{\left< #1 \right|} 

\makeatletter
\renewcommand{\p@subsection}{}
\renewcommand{\p@subsubsection}{}
\makeatother

\usepackage{enumitem}
\setlist{nolistsep}

\def\CO{{\cal O}}

\begin{document}

\title{\large OPE convergence in non-relativistic conformal field theories}

\author{Walter D. Goldberger}
\affiliation{Department of Physics, Yale University, New Haven, CT 06511}
\author{Zuhair U. Khandker}
\affiliation{Physics Department, Boston University, Boston, MA 02215}
\author{Siddharth Prabhu}
\affiliation{Department of Physics, Yale University, New Haven, CT 06511}

\begin{abstract}
Motivated by applications to the study of ultracold atomic gases near the unitarity limit, we investigate the structure of the operator product expansion (OPE) in non-relativistic conformal field theories (NRCFTs).   The main tool used in our analysis is the representation theory of charged (i.e. non-zero particle number) operators in the NRCFT, in particular the mapping between operators and states in a non-relativistic ``radial quantization'' Hilbert space.   Our results include:   a determination of the OPE coefficients of descendant operators in terms of those of the underlying primary state,  a demonstration of convergence of the (imaginary time) OPE in certain kinematic limits, and an estimate of the decay rate of the OPE tail inside matrix elements which, as in relativistic CFTs, depends exponentially on operator dimensions.   To illustrate our results we consider several examples, including a strongly interacting field theory of bosons tuned to the unitarity limit, as well as a class of holographic models.  Given the similarity with known statements about the OPE in $SO(2,d)$ invariant field theories, our results suggest the existence of a bootstrap approach to constraining NRCFTs, with applications to bound state spectra and interactions.   We briefly comment on a possible implementation of this non-relativistic conformal bootstrap program.

\end{abstract}

\maketitle


\newpage

\section{Introduction}
\label{sec:intro}

The theory of non-relativistic particles in the unitarity limit, with $S$-wave scattering length $|a|\rightarrow\infty$, exhibits non-relativistic conformal symmetry~\cite{Mehen:1999nd}. This group of symmetries, called the Schrodinger group~\cite{Hagen:1972pd,Niederer:1972zz} (see also~\cite{Jackiw:1990mb,*Jackiw:1992ih}), is the maximal kinematic invariance group of the free Schrodinger equation, in the same way that the relativistic conformal group is the corresponding invariance group of the free massless Klein-Gordon equation. The Schrodinger algebra includes non-relativistic analogs of scale and special conformal transformations and can be obtained as a non-relativistic limit of the conformal group with particle-number conservation~\cite{Niederer:1974xj}. We refer to Schrodinger-symmetric theories as non-relativistic conformal field theories (NRCFTs) in distinction to their relativistic counterparts (CFTs).

The motivation for studying NRCFTs is twofold. On the one hand, there are several examples of naturally occurring (approximate) NRCFTs. These include few-nucleon systems like the deuteron~\cite{Kaplan:1998tg,*Kaplan:1998we} as well as several atomic systems (\emph{e.g.}, $^{85}$Rb~\cite{PhysRevLett.81.5109}, $^{133}$Cs~\cite{PhysRevLett.85.2717,*PhysRevLett.85.2721}, and $^{39}$K~\cite{PhysRevLett.88.173201}), all of which are characterized by an accidentally large scattering length. More recently, interest in NRCFTs has stemmed from experimentally-tunable cold-atom systems. These are typically cold, dilute Fermi or Bose gases where experimental manipulation of a Feshbach resonance allows one to freely tune the $S$-wave scattering length of the constituent atoms~\cite{PhysRevLett.92.040403,PhysRevLett.92.120403}. In the case of fermionic atoms with two spin states, adjusting the scattering length interpolates the system between the regime of BCS superfluidity at $a^{-1}\sim -\infty$ and BEC superfluidity at $a^{-1}\sim +\infty$, with the unitarity limit $a^{-1}\sim 0$ being the midpoint of this crossover~\cite{Eagles:1969zz,*Leggett1,*Leggett2,*Nozieres}. As such, the unitarity limit and non-relativistic conformal symmetry in general play an essential role in BCS-BEC crossover physics. (For a comprehensive theoretical review of the BCS-BEC crossover and the unitary Fermi gas, see~\cite{ZwergerBook}.)

A major obstacle to theoretical calculations in the unitarity regime in $d=3$ spatial dimensions is that it is strongly coupled, so standard methods from many-body physics are not adequate.    In recent years, it has become apparent that a familiar tool from quantum field theory, namely the operator product expansion (OPE), can be exploited to obtain analytical predictions even in the strongly interacting unitary regime.   In any quantum field theory, the OPE is the statement that an operator product $A(x)B(0)$, in the limit $x\rightarrow 0$, can be expanded in a basis of local operators:
\begin{equation}
\lim_{x\rightarrow 0} A(x)B(0) \sim \sum_\alpha f_\alpha(x) \mathcal{O}_\alpha(0).
\label{OPEGeneral}
\end{equation}
The sum on the right is over all operators allowed by symmetries, and the Wilson coefficients $f_\alpha(x)$ are calculable if the theory is weakly coupled at short distances by evaluating matrix elements on both sides of the OPE relation. In general the OPE is an expansion in the scaling dimension of operators, with higher-dimension operators systematically suppressed.

The utility of the OPE in the unitary regime is that the expansion is dominated by the lowest lying dimension operators, for which the Wilson coefficients can be obtained exactly, in closed analytical form, by evaluating relatively simple one- and two-body matrix elements.    The operator relations obtained this way are \emph{universal}, because they hold true not just inside few-body matrix elements, but also when evaluated inside an arbitrarily complicated many-particle state.   Once a particular many-body state is specified, these operator expansions imply relations between physical observables, namely, the operator expectation values evaluated in the specified state.   This paradigm for applying the OPE to the unitarity regime was initiated in~\cite{Braaten:2008uh}, and further developed in~\cite{Braaten:2008bi,Braaten:2011sz,Barth20112544,PhysRevB.87.235125,Son:2010kq,Goldberger:2010fr,PhysRevA.84.043603,Goldberger:2011hh,Braaten:2013eya}.  For a review of some of these results, see ref.~\cite{Braaten:2010if}.   

\indent These applications motivate a better understanding of the general properties satisfied by the OPE in any NRCFT.    While NRCFT correlation functions have been studied previously (\emph{e.g.},~\cite{Bergman:1991hf,Henkel:1993sg,Henkel:2003pu,Nishida:2007pj,PhysRevA.78.013614,
Volovich:2009yh,Fuertes:2009ex,Leigh:2009eb,Barnes:2010ev,Jackiw:2012ur,Duval:1994pw,*Duval:1994vh,*Duval:2008jg,*Duval:2009vt,Bagchi:2009pe}), less is known about the general structure of the OPE. This is in sharp contrast to the case of CFTs, where the OPE is known to satisfy powerful constraints as a result of conformal symmetry. In particular, the OPE breaks up into conformal multiplets consisting of primary operators of definite scaling dimension and their descendants, obtained by taking spacetime derivatives.   Furthermore, in a CFT the OPE converges~\cite{Luscher:1975js,Mack:1976pa} and does so exponentially fast~\cite{Pappadopulo:2012jk}. Convergence means that if the OPE is used to expand a correlator, the resulting series has finite radius of convergence, determined by the nearest operator insertion.  Exponential convergence means that if the OPE series for a correlation function is truncated at scaling dimension $\Delta_*$ ($\gg 1$), then the error due to the truncation is suppressed by $e^{-\beta\Delta_*}$, where $\beta$ depends on the coordinates of the operators in the correlator.    

In this paper, we analyze the general properties of the OPE in Schrodinger invariant theories.   We find that the OPE in NRCFTs has several properties in common with its relativistic counterpart.   In sec.~\ref{sec:prelim}, we work out the local properties of the OPE, in Galilean coordinates, for NRCFT primary operators ${\cal O}$, defined to have non-zero particle number $N_{\cal O}$ and scaling dimension $\Delta_{\cal O}$.   As in the relativistic case, the OPE for products with net particle number organizes itself into primary operators and descendants, with the descendants' Wilson coefficients determined by those of the primary.   However, unlike CFTs, the Wilson coefficient for the primary operator depends on a function not fixed by the symmetries.   

Starting in sec.~\ref{sec:soc}, we discuss correlation functions and the OPE from a global viewpoint, by working in terms of a set of ``oscillator'' coordinates whose Hamiltonian plays a role similar to the radial quantization Hamiltonian in CFTs.    These coordinates make manifest the state-operator correspondence between  $N_{\cal O}\neq 0$ primaries and the bound state spectrum of the oscillator Hamiltonian, first discussed in~\cite{Nishida:2007pj}.   We use this correspondence, in sec.~\ref{sec:conv}, to provide arguments for the convergence of the OPE, and to give a simple bound on the convergence rate, finding it to be exponential in the cutoff $\Delta_{\star}$ as in the CFT case.   Several explicit examples, analyzed in sec.~\ref{sec:3PF}, provide consistency checks of our results.   Unfortunately, our general results only apply in the sector of operators with $N_{\cal O}\neq 0$, since the decomposition of operators in terms of primaries and descendants, which is crucial to our analysis, is known to break down for the case $N_{\cal O}=0$, see~\cite{Bekaert:2011qd}.

Even though our results cannot be applied to the important case of conserved currents or the stress tensor, they suggest the existence of a version of the conformal bootstrap program~\cite{Ferrara:1973yt,Polyakov:1974gs,Belavin:1984vu} and its recent resurgence in higher dimensions, initiated in~\cite{Rattazzi:2008pe}, for NRCFTs.   Such a non-relativistic bootstrap could have physical applications to constraining few- and many-body bound state spectra and interactions of unitary atoms in a harmonic trap.   We briefly comment on a possible implementation of this program in the conclusions, sec.~\ref{sec:conclusion}.


\section{OPE structure in the Galilean frame}
\label{sec:prelim}

In this paper we consider non-relativistic conformal field theories (NRCFTs), which are invariant not only under Galilean transformations acting on space and time, but also under non-relativistic conformal transformations.   The symmetry generators consist of the (extended) Galilean algebra with generators of space and time translations  $P_i$ ($i=1,\cdots, d$) and $H$, spatial rotations  $M_{ij}=-M_{ji}$,  boosts $K_i$, and $U(1)$ particle number symmetry $N$.  In addition, the conformal transformations consist of dilatations $D$ and special conformal transformations $C$ acting on the coordinates.   The algebra spanned by these generators is the maximal symmetry of the free Schrodinger equation, so it is usually referred to as the Schrodinger algebra (unless otherwise noted, we leave the dimensionality $d$ of space unspecified).   

In this section, we derive constraints on the OPE in the coordinate system $({\vx},t)$ naturally defined by the action of the Galilean generators acting on the origin (the fixed point of rotations).   Acting on these \emph{Galilean coordinates}, we have  $H:(\vx,t)\rightarrow (\vx,t+a)$, $\vec P:(\vx,t)\rightarrow(\vx+{\vec a},t)$, and boosts  ${\vec K}:(\vx,t)\rightarrow (\vx + t {\vec v},t)$, while rotations act linearly $({\vx},t)\rightarrow (R\vx,t)$.    Finite scale transformations generated by $D$ act as
\begin{equation}
D:  (\vx,t)\rightarrow(\lambda \vx,\lambda^{2}t),
\end{equation}
while the action of a finite special conformal transformation,
\begin{equation}
C:  (\vx,t)\rightarrow\left(\frac{\vx}{1+at},\frac{t}{1+at}\right),
\end{equation}
can be factored as $IHI$, with $I$ an \emph{inversion}
\begin{equation}
I:  (\vx,t)\rightarrow\left(\frac{\vx}{t},\frac{1}{t}\right).
\end{equation}
In particular, the generators $H,D,C$ span an $SL(2,\mathbb{R})$ subgroup which acts on the time coordinate as
\begin{equation}
t\rightarrow {\frac{a t + b}{c t + d}},
\end{equation}
with $a d - b c=1$.

The commutation relations involving the generators $(H,P_i,M_{ij},K_i,D,C,N)$ are given by the standard Galilean algebra, as well as 
\begin{equation}
\left[N,\text{any}\right]=0,
\end{equation}
and 
\begin{align}
&[D,P_{i}]=iP_{i},&  \qquad &[D,K_{i}]=-iK_{i},&  \qquad  &[D,C]=-2iC,&   \qquad &[D,H]=2iH, & \nonumber \\
&[H,K_{i}]=-iP_{i},&\qquad  &[H,C]=-iD,& \qquad  &[C,P_{i}]=iK_{i},& \qquad &[K_{i},P_{j}]=i\delta_{ij}N.&
\end{align}
See Appendix~\ref{sec:comm_relns} for the full set of commutation relations.

\subsection{Preliminaries:   Local operators, representations, and correlation functions}

As in the case of relativistic conformal field theories (CFTs), the local operators of a NRCFT, defined such that
\begin{eqnarray}
\label{eq:tt}
[H,{\cal O}(x)] &=& -i \partial_t {\cal O}(x), \\
\label{eq:st}
{}  [P_i,{\cal O}(x)] &=& i \partial_i {\cal O}(x), 
\end{eqnarray}
 can be expanded into a set of ``lowest weight'' primary operators with definite scaling dimension plus their \emph{descendants}, obtained by taking space and time derivatives of the primary operators~\cite{Nishida:2007pj}.   Each primary operator and its tower of descendants fills out a representation of the Schrodinger group.   The representations are constructed by translating the operators to the origin $t={\vec x}=0$ and decomposing into irreducible representations of the stability group of that point, which consists of the subgroup generated by rotations, Galilean boosts, dilatations, and special conformal transformations.    

Thus we decompose the operators at the origin into eigenstates of particle number,
\begin{equation}
\left[N,\mathcal{O}(0)\right] = N_{\mathcal{O}}\mathcal{O}(0),
\end{equation}
where $N_{\mathcal{O}}$ is restricted to be integer, and of dilatations,
\begin{equation}
[D,\mathcal{O}(0)] = i\Delta_{\mathcal{O}}\mathcal{O}(0),
\end{equation}
where $\Delta_{\cal O}$ is the \emph{scaling dimension}. There is a unitarity bound~\cite{Nishida:2010tm} for the scaling dimension, $\Delta_{\cal O}\geq d/2$, as we will review in sec.~\ref{sec:soc}.    
Note that as a consequence of the Schrodinger algebra, the operators $[{ K_i},{\cal O}(0)]$ and $[{C},{\cal O}(0)]$ have scaling dimensions $\Delta_{\cal O}-1$ and $\Delta_{\cal O}-2$.   The unitarity bound then implies that among the eigenstates of $(D,N)$ there exists a set of lowest weight \emph{primary operators} with $\Delta_{\cal O}\geq d/2$ such that
\begin{equation}
\left[{K}_i,\mathcal{O}(0)\right]=\left[C,\mathcal{O}(0)\right]=0.
\end{equation}    

From Eqs.~(\ref{eq:tt}), (\ref{eq:st}), an operator ${\cal O}(x)$ at any point away from the origin is given by 
\begin{equation}
\mathcal{O}(x)=e^{ix\cdot P}\mathcal{O}(0)e^{-i x\cdot P}= {\cal O}(0) + [i x\cdot P,{\cal O}(0)]+{1\over 2!} [i x\cdot P,[i x\cdot P,{\cal O}(0)]]+\cdots,
\end{equation}
where $x\cdot P = t H - {\vec x}\cdot {\vec P}$. Thus, if ${\cal O}(0)$ is a primary operator, $\mathcal{O}(x)$  is a linear combination of this primary and its descendants.  From the algebra, it follows that $[P_{i},\mathcal{O}(0)]$ and $[H,{\cal O}(0)]$ have scaling dimensions $\Delta_{\cal O}+1$ and $\Delta_{\cal O}+2$ respectively.   Therefore an irreducible unitary representation of the Schrodinger group consists of a primary operator of lowest dimension $\Delta_{\cal O}$ together with an infinite tower of descendant states.  It is straightforward to see that away from the origin, in addition to obeying Eqs.~(\ref{eq:tt}), (\ref{eq:st}), the primary operator ${\cal O}(x)$ has the following commutators with the Schrodinger group generators, 
\begin{eqnarray}
\left[N,\mathcal{O}(x)\right] &=& N_{\mathcal{O}}\mathcal{O}(x)\\
  \left[D,\mathcal{O}(x)\right]&=&i\left(2t\partial_t+x_i\partial_i+\Delta_\mathcal{O}\right)\mathcal{O}(x),\\
  \left[C,\mathcal{O}(x)\right] &=& \left(-it^2\partial_t-itx_i\partial_i-it\Delta_\mathcal{O}+{N_{\cal O}\over 2}{\vx}^2\right) \mathcal{O}(x) \label{CAction},\\
  \left[K_i,\mathcal{O}(x)\right] &=& \left(-it\partial_i+N_\mathcal{O}x_i\right) \mathcal{O}(x) \label{KAction}.
  \end{eqnarray}

It is important to note that the representation structure just described only makes sense for operators with nonzero particle number~\cite{Bekaert:2011qd,Perroud:1977qh}.  In the sector with $N_{\cal O}=0$, the Galilean algebra implies $[{K}_i,{P}_j]=0$, and the decomposition of operators into primaries and descendants breaks down:   in particular, if ${\cal O}(0)$  is a primary operator, so is the descendant $[{ P}_i,{\cal O}(0)]$.   Because the decomposition into primary and descendant operators is crucial for obtaining our general results below, we focus on the case of operators $N_{\cal O}\neq0$.    This unfortunately limits the applications of the results presented in this paper, as we cannot make statements about the OPE of $N_{\cal O}=0$ operators such as the stress tensor $T^{ij}$ and the particle current $J^i$, which are relevant for hydrodynamic transport.   In this case, one must resort instead to explicit calculation of the OPE within a given model, as was done in refs.~\cite{Braaten:2008bi,Braaten:2011sz,Barth20112544,PhysRevB.87.235125,Son:2010kq,Goldberger:2010fr,Goldberger:2011hh,PhysRevA.84.043603,Braaten:2013eya}.

\subsubsection{Correlation functions}

In relativistic CFTs, conformal invariance uniquely fixes the coordinate dependence of primary operator two- and three-point correlators.  On the other hand, the four-point function depends on two possible conformally invariant cross ratios involving the coordinates, and it is therefore not fully determined by conformal kinematics alone.   In the non-relativistic case, there exists a three-point invariant cross ratio, meaning that only the two-point function in the sector of operators with  $N_{\cal O}\neq 0$ is fixed by symmetry.   Nevertheless, it is possible to make some general statements about the structure of the $n$-point correlators, and we review the relevant results in this section for the case of rotational scalar primary operators.

Consider correlation functions of primary operators $\phi_i(x)$, defined as 
\begin{equation}
G_{n}\left(1,2,\ldots n\right)\equiv\bra{0}\phi_{1}(x_1)\cdots \phi_n(x_n)\ket{0},
\end{equation}
where the state $\ket{0}$ is the trivial vacuum of the NRCFT, annihilated by all the symmetry generators.   In particular, 
\begin{equation}
H\ket{0} = N\ket{0} = 0.
\end{equation}
Since $\phi_i(x)$ has definite particle number $N_i$, the correlator is non-zero only for  $\sum_i N_{i}=0$.   By scale, translation and rotational invariance, the two-point function is of the form
\begin{equation}
G_2(x_1,x_2) = \bra{0} \phi_1(x_1) \phi_2(x_2)\ket{0} = |t_{12}|^{-(\Delta_1+\Delta_2)/2} f({\vec x}_{12}^2/t_{12}),
\end{equation}   
where $x_{ij}=x_i-x_j$.  The action of Galilean boosts then fixes the function $f(z)= c_{12}e^{-i N_1 z/2}$, for some constant $c_{12}$ while special conformal transformations require that $c_{12}=0$ unless $\Delta_1=\Delta_2$.    Thus the two-point function is fixed to be~\cite{Henkel:1993sg}
\begin{equation}
G_{2}(x_1,x_2)=\delta_{\Delta_{1},\Delta_{2}}\,c\,  t^{-\Delta_{1}}_{12}\exp\left\{ -i\frac{N_{1}}{2}\frac{{\vx}^{2}_{12}}{t_{12}}\right\},
\end{equation}
where the branch of $t_{12}^{-\Delta}$ is fixed by choosing a suitable $i\epsilon$ prescription.     Without loss of generality, we can choose our operator basis such that $\phi_2(x)=\phi_1^\dagger(x)$ for $N_{\phi_1}<0$ and thus the general two-point function is 
\begin{equation}
G_{2}(x_1,x_2)= \langle 0|\phi_1(x_1)\phi_1^\dagger(x_2)|0\rangle=  c  t^{-\Delta_{1}}_{12}\exp\left\{ -i\frac{N_{1}}{2}\frac{{\vx}^{2}_{12}}{t_{12}}\right\}.
\label{eq:2pf} 
\end{equation}

For $n>2$ operator insertions, the correlator can depend on a number of conformally invariant cross ratios of the coordinates.  In the non-relativistic case, there are two kinds of invariants.  First, we have cross-ratios that are analogous to those that arise in CFTs, but now only involving
time, of the form 
\begin{equation}
\frac{t_{ij}t_{kl}}{t_{il}t_{jk}},
\end{equation}
with $i,j,k,l=1,\cdots, n$.  The second type involve both spatial and time coordinates and can be parametrized as
\begin{equation}
v_{ijn}\equiv\frac{1}{2}\frac{\left(\vec{x}_{in}t_{jn}-\vec{x}_{jn}t_{in}\right)^{2}}{t_{ij}t_{in}t_{jn}}=\frac{1}{2}\left(\frac{{\vx}_{jn}^{2}}{t_{jn}}+\frac{{\vx}_{ij}^{2}}{t_{ij}}-\frac{{\vx}_{in}^{2}}{t_{in}}\right)\qquad i<j<n .
\label{vij}
\end{equation}
The most general three-~\cite{Henkel:1993sg} and four-~\cite{Volovich:2009yh} point correlators of scalar primary operators consistent with the symmetries are of the form\begin{eqnarray}
G_{3}(1,2,3) &=&\left[\exp\left\{ -i\frac{N_{1}}{2}\frac{{\vx}_{13}^{2}}{t_{13}}-i\frac{N_{2}}{2}\frac{{\vx}_{23}^{2}}{t_{23}}\right\} \prod_{i<j}t_{ij}^{\Delta/2-\Delta_{i}-\Delta_{j}}\right]\times  F\left(v_{123}\right), \label{3pf} \\
G_{4}(1,2,3,4)&=&\left[\exp\left\{ -i\frac{N_{1}}{2}\frac{{\vx}_{14}^{2}}{t_{14}}-i\frac{N_{2}}{2}\frac{{\vx}_{24}^{2}}{t_{24}}-i\frac{N_{3}}{2}\frac{{\vx}_{34}^{2}}{t_{34}}\right\} \prod_{i<j}t_{ij}^{\Delta/6-\left(\Delta_{i}+\Delta_{j}\right)/2}\right]  \nonumber \\
&& {}\times H\left(v_{124},v_{134},v_{234},\frac{t_{12}t_{34}}{t_{13}t_{24}} \right),
\end{eqnarray}
where $\Delta\equiv\sum_i\Delta_i$.  In these equations, conformal kinematics fixes the dependence on coordinates up to model-dependent scalar functions $F$, $H$ of the invariants. 

\subsection{OPE constraints}

\label{sec:constraints}

Given the above facts about NRCFT operators, we may now place constraints on the form of their OPE.   In any quantum field theory, the OPE is the statement that the product of operators at nearly coincident points has an expansion in local operators,
\begin{equation}
\lim_{x\rightarrow 0} \phi_1(x) \phi_2(0)\sim \sum_\alpha c_\alpha(x) {\cal O}(0).
\end{equation}
This equation should be regarded as an operator equation, whose meaning is that when inserted into an arbitrary matrix element, the RHS is an asymptotic expansion for the LHS.   

In CFTs, it is possible to make more explicit statements regarding the structure of the OPE.   In particular, scale invariance implies that the Wilson coefficients are given up to constants by $c_\alpha(x)\sim \left(x^2\right)^{(\Delta_\alpha-\Delta_{\phi_1}-\Delta_{\phi_2})/2}$.  Using special conformal symmetry, for any given primary operator ${\cal O}$ appearing in the OPE, the Wilson coefficients of descendants $\partial_{\mu_1}\cdots \partial_{\mu_n} {\cal O}(0)$ are fixed entirely by the conformal algebra in terms of the coefficient of the primary ${\cal O}$~\cite{Ferrara:1971zy,*Ferrara:1971zz,*Ferrara:1972cq}.   A relatively straightforward way of obtaining these constraints is to compute commutators of the conformal generators on both sides of the OPE.   Using the known conformal transformation properties of the operators $\phi_1(x)\phi_2(0)$ and of ${\cal O}(x)$, these commutators yield recurrence formulae that relate the coefficients $c_\alpha(x)$ to those of the primary ${\cal O}(0)$.

In NRCFTs, it is similarly useful to decompose the OPE into contributions from different primaries together with their descendant towers.    In this case, scale invariance does not completely fix the functional form of the Wilson coefficients, and the OPE takes a more general form, with a slight distinction between the cases of $d>1$ and $d=1$ spatial dimensions. We discuss the case $d=1$ in sec.~\ref{sec:1d} below. For now, we focus on $d>1$, in which case the most general structure of the OPE is
\begin{equation}
\lim_{x\rightarrow 0} \phi_1(x)\phi_2(0)\thicksim\sum_{\cal O}\sum_{r,s,q}f^{\cal O}_{r,s,q}(\vx,t)\mathcal{O}^{r,s,q}.
\label{OPE1}
\end{equation}
Here, the outer sum is over primary operators ${\cal O}$ with particle number $N_{\cal O}=N_{\phi_1}+N_{\phi_2}$, which is by assumption non-zero.   The descendants ${\cal O}^{r,s,q}$, with dimension $\Delta^{\cal O}_{rsq}=\Delta_{\cal O}+ 2(s+q)$, are obtained by acting on ${\cal O}$ with $r$ spatial derivatives, $s$ powers of the Laplacian $\nabla^2 = \partial_i \partial_i$, and $q$ time derivatives $\partial_t$,
\begin{equation}
\mathcal{O}^{r,s,q}\equiv x^{j_{1}}\cdots x^{j_{r}} (i\partial_{j_{1}})\cdots (i\partial_{j_{r}})(-\nabla^{2})^{s}(-i\partial_t)^{q}\CO(0).
\label{Odesc}
\end{equation}
Covariance under translations, rotations and dilatations suggests that the Wilson coefficients $f^{\cal O}_{r,s,q}(\vx,t)$ be parametrized as
\begin{equation}
f^{\cal O}_{r,s,q}(x)=\frac{1}{r!s!q!}\frac{1}{\left(2N_{\mathcal{O}}\right)^{s}}t^{-\frac{1}{2}\Delta_{\phi_1\phi_2,\mathcal{O}}+q+s}e^{-iN_{\phi_1}{\vx}^{2}/(2t)}c_{r,s,q}\left(z\equiv-N_{\mathcal{O}}\frac{\vx^{2}}{2t}\right),
\label{eq:cdef}
\end{equation}
for some unknown functions $c_{r,s,q}(z)$.   In this equation we have defined
\begin{equation}
\Delta_{\alpha\beta,\gamma} \equiv \Delta_\alpha + \Delta_\beta - \Delta_\gamma.
\end{equation}
The factor of $e^{-iN_{\phi_1}{\vx}^{2}/(2t)}$ has been inserted in order to simplify the action of Galilean boosts.    The precise form of the limit $x\rightarrow 0$ is not important in this section, but will matter when we discuss OPE convergence in sec.~\ref{sec:conv}.

We now compute commutators of both sides of the OPE with the Schrodinger generators.   In writing Eq.~(\ref{eq:cdef}), we have already accounted for the constraints imposed by the commutators with $M_{ij}$, ${\vec P}_i$, $H$ and $D$.   The remaining commutators with ${\vec K}_i$ and $C$ yield relations among the functions $c_{r,s,q}(z)$ for different values of the integers $r,s,q$.    To find these relations, we use on the LHS of the OPE,
\begin{eqnarray}
 [K_i,\phi_1(x)\phi_2(0)] &=& [K_i,\phi_1(x)]\phi_2(0) =\left(-it\partial_i+N_{\phi_1} x_i\right)\left[\phi_1(x)\phi_2(0)\right],\\
{} [C,\phi_1(x)\phi_2(0)] &=& [C,\phi_1(x)]\phi_2(0) = \left(-it^2\partial_t-itx_i\partial_i-it\Delta_{\phi_1}+{N_{\phi_1}\over 2}{\vx}^2\right)\left[\phi_1(x)\phi_2(0)\right].
\end{eqnarray}
On the RHS of the OPE, the action of the commutators is, from Eqs.~(\ref{CAction})-(\ref{KAction}), 
\begin{eqnarray}
\left[K_{i},\mathcal{O}^{r,s,q}\right] &=& ir N_{\mathcal{O}} x_{i}\mathcal{O}^{r-1,s,q} + 2isN_{\mathcal{O}}  P_{i}\mathcal{O}^{r,s-1,q}+iq P_{i}\mathcal{O}^{r,s,q-1}, \label{KOrsq} \\
\left[C,\mathcal{O}^{r,s,q}\right] &=&
-\frac{1}{2}N_{\mathcal{O}}r\left(r-1\right){\vx}^{2}\mathcal{O}^{r-2,s,q}-2N_{\mathcal{O}}s\left(d/2+r+s-1\right)\mathcal{O}^{r,s-1,q} \nonumber \\
&& -q\left(\Delta_{\mathcal{O}}+r+2s+q-1\right)\mathcal{O}^{r,s,q-1}, 
\label{COrsq-1}
\end{eqnarray}
where in Eq.~(\ref{KOrsq}), $P_{i}\mathcal{O}^{r,s,q} \equiv x^{j_{1}}\cdots x^{j_{r}}(i\partial_{i})\mathcal{O}_{j_{1}\cdots j_{r}}^{s,q}(0)$.   Equating the action of $[K_i,\cdot]$ on both sides of the OPE yields two constraints, which are
\begin{equation}
c_{r+1,s,q}(z)=\partial_{z}c_{r,s,q}(z) \label{cRecursion1},
\end{equation}
and 
\begin{equation}
c_{r+1,s,q}(z)+c_{r,s+1,q}(z)+c_{r,s,q+1}(z)=0. \label{cRecursion2}
\end{equation}
On the other hand, special conformal transformations yield the sole constraint
\begin{eqnarray}
\nonumber
\left(z\partial_z  + q+r+s+{1\over 2}\Delta_{\phi_1{\cal O},\phi_2}\right)c_{r,s,q}(z) &=& iz c_{r+2,s,q}(z) - i\left({d\over 2}+r+s\right)c_{r,s+1,q}(z) \\
& & {} - i\left(\Delta_{\cal O} + r + 2s +q\right) c_{r,s,q+1}(z).
\end{eqnarray}

These three equations comprise a set of recursion relations for the functions $c_{r,s,q}(z)$.  Given $c_{r,s,q}(z)$, Eq.~(\ref{cRecursion1}) fixes $c_{r+1,s,q}(z)$, while the remaining constraints can be re-written as
\begin{eqnarray}
c_{r,s,q+1}(z) &=&\frac{\left[z\partial_{z}^{2}+\left(iz+\frac{d}{2}+r+s\right)\partial_{z}+i\left({1\over 2}\Delta_{\phi_1\mathcal{O},\phi_2}+q+r+s\right)\right]c_{r,s,q}(z)}{\left(\Delta_{\mathcal{O}}-\frac{d}{2}+q+s\right)} \label{cRecursion3},\\
\nonumber\\
c_{r,s+1,q}(z) &=& -c_{r+1,s,q}(z)-c_{r,s,q+1}(z) \nonumber\\
&=& -\frac{\left[z\partial_{z}^{2}+\left(iz+\Delta_{\cal O}+r+2s+q\right)\partial_{z}+i\left({1\over 2}\Delta_{\phi_1\mathcal{O},\phi_2}+q+r+s\right)\right]c_{r,s,q}(z)}{\left(\Delta_{\mathcal{O}}-\frac{d}{2}+q+s\right)}.
\end{eqnarray}
Thus $c_{r,s,q}(z)$ determines $c_{r+1,s,q}(z)$, $c_{r,s+1,q}(z)$, $c_{r,s,q+1}(z)$, and by induction all Wilson coefficients $c_{r,s,q}(z)$ are determined by $c_{0,0,0}(z)$.     

Note, however, that $c_{0,0,0}(z)$ is not determined by conformal kinematics alone.   This is related to the fact that the three-point function, Eq.~(\ref{3pf}), depends on an a priori unknown function $F(v)$ of the three-point conformal invariant.   Indeed, inserting both sides of the OPE into the correlation function $\bra{0}|{\cal O}^\dagger\left[\cdots\right]|\ket{0}$ yields that 
\begin{equation}
c_{0,0,0}(z) = e^{-i{N_{\phi_1}\over N_{\cal O}} z}F\left(v=-{z\over N_{\cal O}}\right) / c_\mathcal{O},
\end{equation}
where $c_\mathcal{O}$ is the constant appearing in the two-point function of $\mathcal{O}$ in Eq.(\ref{eq:2pf}). We stress that in obtaining these results, it is essential that $\mathcal{O}(x)$ has non-zero number charge. For operators with $N_{\cal O}=0$, the primary/descendant structure is more complicated, and the categorization of descendants in Eq.~(\ref{Odesc}) ceases to be valid.  The recursion relations above only apply to $SO(d)$ scalar primaries, but can be straightforwardly generalized to higher spin representations.

\indent As a simple example of these equations, we consider the field theory of a free boson $\phi(x)$,
\begin{equation}
{\cal L} = \phi^\dagger \left( i\partial_t + {1\over  2}{\nabla^2}\right)\phi.
\label{eq:free}
\end{equation}
Because in free field theory $\Delta_{\phi^2}=2 \Delta_\phi=d$, charge conservation implies that the OPE is non-singular, and can therefore be obtained by simply Taylor expanding the fields,
\begin{equation}
\phi(x)\phi(0)\sim\sum_{p=0}^{\infty}\sum_{q=0}^{\infty}\frac{1}{p!q!}t^{p}x^{i_{1}}\cdots x^{i_{q}}\left(\partial_{t}^{p}\partial_{i_{1}}\cdots\partial_{i_{q}}\phi(0)\right)\phi(0).
\label{FreeOPE}
\end{equation}
Note that both sides of this equation are automatically normal ordered.     

The OPE written in this way disguises the decomposition of the OPE into distinct Schrodinger representations.    In order to make contact with our formalism, the RHS of this equation has to be re-expressed in a basis of primary operators and their descendants.    For instance, the two lowest-dimension primary operators appearing in the OPE are
\begin{eqnarray}
\phi^{2},\qquad  && \Delta_{\phi^{2}}=d \\
J_{ij}\equiv\phi\overleftrightarrow{\partial_{i}}\overleftrightarrow{\partial_{j}}\phi-\frac{1}{d}\delta_{ij}\phi\overleftrightarrow{\partial}\cdot\overleftrightarrow{\partial}\phi,\qquad && \Delta_{J_{ij}}=d+2
\end{eqnarray}
\noindent Subtracting out the trace piece in $J_{ij}$ is necessary to ensure orthogonality of the operator basis, i.e. $\left\langle \phi^{\dagger2}J_{ij}\right\rangle=0$. The derivatives in Eq.~(\ref{FreeOPE}) can be rearranged using the chain rule and the free theory equation of motion, $\nabla^{2}\phi=(2iN_\phi)\partial_{t}\phi$,
to get
\begin{eqnarray}
\phi(x)\phi(0)&\sim& \left[1+\frac{1}{2}x^{i}\partial_{i}+\frac{1}{8}x^{i}x^{j}\partial_{i}\partial_{j}-\frac{{\vx}^{2}}{8d}\nabla^{2}+\left(\frac{t}{2}+\frac{iN_\phi}{2d}{\vec x}^{2}\right)\partial_{t}+\cdots\right]\phi^{2}(0) \nonumber \\
&& +\left[\frac{1}{8}x^{i}x^{j}+\cdots\right]J_{ij}(0)+\text{other reps.},
\label{FreeOPE2}
\end{eqnarray}
which now makes the distinct representations manifest.   In particular, $c_{0,0,0}^{(\phi^{2})}(z)=e^{-iz/2}$ for the $\phi^{2}$
representation, and the functions $c_{r,s,q}^{(\phi^{2})}(z)$ appearing in front of the $\phi^{2}$ descendants are fixed by the recursion relations.   
%

\subsubsection{Constraints in d=1}
\label{sec:1d}

In one spatial dimension, there is no need to distinguish the Laplacian in the OPE and Eqs.~(\ref{OPE1}-\ref{eq:cdef}) take the simpler form,
\begin{equation}
\lim_{x\rightarrow 0}\phi_{1}(x)\phi_{2}(0)\sim\sum_{\mathcal{O}}\sum_{r,q}f_{r,q}^{\mathcal{O}}(x,t)\mathcal{O}^{r,q},
\end{equation}
where now
\begin{equation}
\mathcal{O}^{r,q}\equiv x^{r}\left(i\partial_{x}\right)^{r}\left(-i\partial_{t}\right)^{q}\mathcal{O}(0),
\end{equation}
and
\begin{equation}
f_{r,q}^{\mathcal{O}}(x)=\frac{1}{r!q!}t^{-\frac{1}{2}\Delta_{\phi_{1}\phi_{2},\mathcal{O}}+q}e^{-iN_{\phi_{1}}\frac{x^{2}}{2t}}c_{r,q}\left(z=-N_{\mathcal{O}}\frac{x^{2}}{2t}\right).
\end{equation}
The commutators with $K$ and $C$ are given by
\begin{eqnarray}
\left[K,\mathcal{O}^{r,q}\right]&=&iN_{\mathcal{O}}r\mathcal{O}^{r-1,q}+iq\mathcal{O}^{r+1,q-1}, \\
\left[C,\mathcal{O}^{r,q}\right]&=&-N_{\mathcal{O}}\frac{r\left(r-1\right)}{2}\mathcal{O}^{r-2,q}-q\left(\Delta_{\mathcal{O}}+r+q-1\right)\mathcal{O}^{r,q-1},
\end{eqnarray}
and the resulting constraints are, respectively,
\begin{eqnarray}
\left(2z\partial_{z}+r\right)c_{r,q}&=&2zc_{r+1,q}-rc_{r-1,q+1}\label{1dKRecursion} \\
\left[z\partial_{z}+q+r+\frac{1}{2}\Delta_{\phi_{1}\mathcal{O},\phi_{2}}\right]c_{r,q}&=&izc_{r+2,q}-i\left(\Delta_{\mathcal{O}}+r+q\right)c_{r,q+1}\label{1dCRecursion}
\end{eqnarray}
Given $c_{0,0}(z)$, Eq.~(\ref{1dKRecursion}) for $r=0$ determines $c_{1,0}(z)$. The remaining $c_{r,q}(z)$ are determined recursively, for instance, by rewriting the above constraints as
\begin{eqnarray}
c_{r,q+1}&=&\frac{\left[\left(2z\partial_{z}+r+1\right)c_{r+1,q}+2i\left(z\partial_{z}+q+r+\frac{1}{2}\Delta_{\phi_{1}\mathcal{O},\phi_{2}}\right)c_{r,q}\right]}{\left(2\Delta_{\mathcal{O}}+r+2q-1\right)} \\
zc_{r+2,q}&=&\frac{\left[\left(\Delta_{\mathcal{O}}+r+q\right)\left(2z\partial_{z}+r+1\right)c_{r+1,q}+i\left(r+1\right)\left(z\partial_{z}+q+r+\frac{1}{2}\Delta_{\phi_{1}\mathcal{O},\phi_{2}}\right)c_{r,q}\right]}{\left(2\Delta_{\mathcal{O}}+r+2q-1\right)}
\end{eqnarray}

We note that when $z=0$, these recursion relations are easily solved. In this limit, $c_{r,q}(z=0)$ are just constants (this follows from analyticity as will be shown in sec.~\ref{sec:conv}), and Eqs.~(\ref{1dKRecursion}-\ref{1dCRecursion}) become
\begin{equation}
c_{r,q+1}=-c_{r+1,q}=\frac{i\left(q+r+\frac{1}{2}\Delta_{\phi_{1}\mathcal{O},\phi_{2}}\right)}{\left(\Delta_{\mathcal{O}}+r+q\right)}c_{r,q} \hspace{10mm} (z=0),
\end{equation}
whose solution is
\begin{equation}
c_{r,q}=\left(-1\right)^{r}i^{r+q}\frac{\left(\frac{1}{2}\Delta_{\phi_{1}\mathcal{O},\phi_{2}}\right)_{r+q}}{\left(\Delta_{\mathcal{O}}\right)_{r+q}}c_{0,0} \hspace{10mm} (z=0),
\end{equation}
where $(a)_s=\Gamma(a+s)/\Gamma(a)$ is the Pochhammer symbol.


\section{Hilbert space structure and OPE in the oscillator frame}
\label{sec:soc}

\begin{figure}
\centering
\includegraphics[scale=.85]{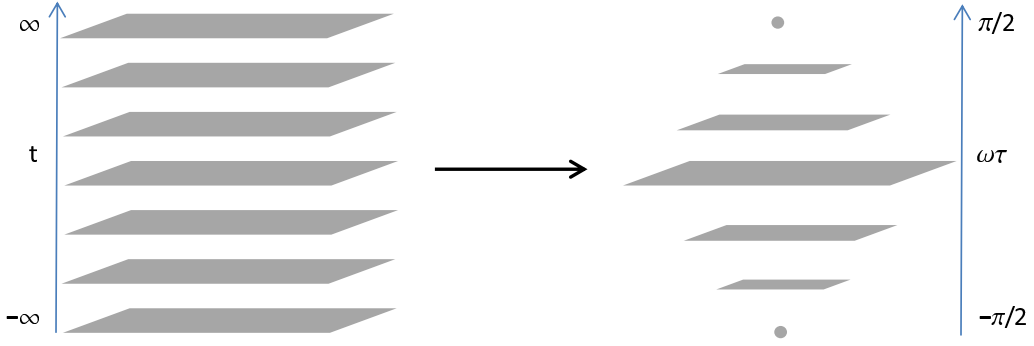}
\caption{Schematic depiction of how time-slices map under the transformation between Galilean and oscillator coordinates.}
\label{fig:osc}
\end{figure}

In this section we discuss the structure of the OPE from the point of view of a coordinate system $y=(\tau,{\vec y})$ which we refer to as the oscillator frame.   These coordinates are related to the Galilean coordinates $(t,\vx)$ by the transformation rules
\begin{equation}
\begin{array}{c}
\omega t=\tan\omega\tau\\
\vec{x}=\vec{y}\sec\omega\tau
\end{array}\longleftrightarrow\begin{array}{c}
\omega\tau=\arctan\omega t\\
\vec{y}=\frac{\vec{x}}{\sqrt{1+\omega^{2}t^{2}}}.
\end{array}
\end{equation}
In particular, the translation $(\tau,{\vec y})\rightarrow(\tau+a,{\vec y})$ is generated by the following linear combination of Schrodinger algebra generators
\begin{equation}
H_{\omega}\equiv H+\omega^{2}C.
\end{equation}
 Geometrically, the map to oscillator coordinates takes constant $t$-slices to constant $\omega\tau$-slices,
after rescaling each slice by a $t$-dependent factor. In particular,
the time slices at $t=-\infty$ and $t=+\infty$ get mapped to a single
point at $\left(\omega\tau=-\frac{\pi}{2},\vec{y}=0\right)$ and $\left(\omega\tau=+\frac{\pi}{2},\vec{y}=0\right)$,
respectively, as depicted in Figure~\ref{fig:osc}.

The oscillator coordinates are best suited for analyzing the operator spectrum of the CFT.    This is because, as reviewed below, the eigenstates of $H_{\omega}$ are in one-to-one correspondence with the set of NRCFT primary operators ${\cal O}$ and their descendants, and thus the spectrum of $H_{\omega}$ coincides with the spectrum of scaling dimensions $\Delta_{\cal O}$~\cite{Nishida:2007pj}.  Thus NRCFT operators generate a Hilbert space, and one can use properties such as completeness and positivity of the norm to constrain their properties.   The oscillator frame can be thought of as the non-relativistic analog of radial quantization in ordinary CFTs, where the spectrum of scaling dimensions is that of the Hamiltonian $D$ that evolves states along the radial direction.    We will use the Hilbert space structure of operators in the oscillator frame to make statements about the convergence of the OPE in sec.~\ref{sec:conv}.

From a more physical perspective, in the case of NRCFTs that arise as many-body quantum field theories with elementary fields $\psi(x)$ whose Galilean frame Lagrangian is of the form
\begin{equation}
{\cal L} = \psi^\dagger \left(i\partial_t + {\nabla^2\over 2 m}\right)\psi + {\cal L}_{int}(\psi,\psi^\dagger),
\end{equation}  
(the interaction ${\cal L}_{int}(\psi,\psi^\dagger)$ is local in time, although not necessarily in space), the special conformal generator is 
\begin{equation}
C={1\over 2}\int d^3{\vec x}\, \vec{x}^2 \psi^\dagger \psi.
\end{equation}
Thus the Hamiltonian $H_\omega$ has a physical interpretation as the Hamiltonian for the NRCFT placed in an external confining harmonic potential.   It follows that the spectrum of $H_\omega$ can be interpreted as the spectrum of $N$-particle bound states in a harmonic trap.    So there is a direct map between experimentally observable properties of the system and the spectrum of primary operators ${\cal O}$.

The fact that there is a kinematic relation between dynamics in the Galilean and oscillator frames was first pointed out, in the case ${\cal L}_{int}=0$ by Niederer~\cite{Niederer:1973tz}.    Generalizing his results to the interacting case, we define a map between primary operators ${\cal O}(x)$ in the Galilean frame and their counterparts ${\tilde{\cal O}}(y)$ in oscillator coordinates by
\begin{equation}
\label{eq:oxtoy}
\tilde{\mathcal{O}}(y)=\left(1+\omega^{2}t^{2}\right)^{{\Delta_{\mathcal{O}}}/{2}}\exp\left[\frac{i}{2}N_{\mathcal{O}}\frac{{\vx}^{2}\omega^{2}t}{1+\omega^{2}t^{2}}\right]\mathcal{O}(x),
\end{equation}
or 
\begin{equation}
\label{eq:oytox}
{\mathcal{O}}(x)=\left[\cos\omega\tau\right]^{{\Delta_{\cal{O}}}} \exp{\left[-{i\over 2} N_{\cal O}\omega{\vec y}^2\tan\omega\tau\right]}\tilde{\mathcal{O}}(y).
\end{equation}
Defining the generators
\begin{eqnarray}
\label{eq:ppm}
\vec{P}_{\pm} &\equiv& \frac{1}{\sqrt{2\omega}}\vec{P}\pm i\sqrt{\frac{\omega}{2}}\vec{K}, \\
\label{eq:lpm}
L_{\pm} &\equiv& \frac{1}{2}\left(\frac{1}{\omega}H-\omega C\pm iD\right),
\end{eqnarray}
it is straightforward to show using Eqs.~(\ref{eq:oxtoy}), (\ref{eq:oytox}) that the operator ${\tilde{\cal O}}(y)$ obeys the commutation relations
\begin{eqnarray}
\label{eq:goo1}
\left[H_{\omega},\tilde{\mathcal{O}}(y)\right] &=& -i\partial_{\tau}\tilde{\mathcal{O}}(y), \\
\label{eq:goo1.5}
\left[\vec{P}_{\pm},\tilde{\mathcal{O}}(y)\right] &=& \frac{i}{\sqrt{2\omega}}e^{\mp i\omega\tau}\left(\vec{\partial}_{y}\pm N_{\mathcal{O}}\omega\vec{y}\right)\tilde{\mathcal{O}}(y), \\
\label{eq:goo2}
\left[L_{\pm},\tilde{\mathcal{O}}(y)\right] &=& \frac{1}{2}e^{\mp2i\omega\tau}\left(-\frac{i}{\omega}\partial_{\tau}\mp\vec{y}\cdot\vec{\partial}_{y}-\left(N_\mathcal{O}\omega {{\vec y}}^{2}\pm\Delta_\mathcal{O}\right)\right)\tilde{\mathcal{O}}(y).
\end{eqnarray}

\subsection{Hilbert space structure in the oscillator frame}

As first pointed out in~\cite{Nishida:2007pj}, a primary operator $\mathcal{O}^{\dagger}(x)$ with $N_{{\cal O}^\dagger}=-N_{\cal O}>0$ and scaling dimension $\Delta_{\cal O}$ defines a state $\left|\mathcal{O}\right\rangle$ given by its action on the Schrodinger invariant vacuum state $|0\rangle$
\begin{equation}
\label{eq:map}
\left|\mathcal{O}\right\rangle \equiv e^{-H/\omega}\mathcal{O}^{\dagger}(0)\left|0\right\rangle,
\end{equation}
with 
\begin{eqnarray}
N\left|\mathcal{O}\right\rangle &=& N_{\mathcal{O}^{\dagger}}\left|\mathcal{O}\right\rangle ,\\
H_{\omega}\left|\mathcal{O}\right\rangle &=& \omega \Delta_{\mathcal{O}}\left|\mathcal{O}\right\rangle.
\end{eqnarray}
These equations follow straightforwardly from the Schrodinger algebra as well as the invariance of the vacuum $|0\rangle$, and imply that, physically, the operator $\left[{\cal O}(t=-i/\omega,0)\right]^\dagger$ translated to imaginary time\footnote{From ${\cal O}(t) = e^{i H t} {\cal O}(0) e^{-i Ht}$, for imaginary time $t$ we have $\left[{\cal O}(t)\right]^\dagger=  e^{i H t^*} {\cal O}^\dagger(0) e^{-i H t^*}= {\cal O}^\dagger(t^*)$ and thus $|{\cal O}\rangle = e^{-H/\omega} {\cal O}^\dagger(0)|0\rangle = \left[{\cal O}(-i/\omega)\right]^\dagger|0\rangle$ while $\langle{\cO}|=\langle 0| {\cal O}(0)e^{-H/\omega}=\langle 0|{\cal O}(-i/\omega)$.}, creates an $N_{\mathcal{O}^\dagger}$-particle bound state with definite energy $E_{\cal O} = \omega \Delta_{\cal O}$ out of the vacuum.     Thus $H_\omega$ plays an analogous role in oscillator coordinates to the dilatation operator $D$ in the Galilean frame.

Similarly the fact that in the Galilean frame the generators ($\vec{P}$, $H$) raise the dimension of an operator while ($\vec{K}$, $C$) lower it corresponds to the fact that in the oscillator frame the generators ${\vec P}_{\pm}$ and $L_\pm$ raise and lower the energy by $\omega$ or $2\omega$,
\begin{eqnarray}
[H_\omega,{\vec P}_\pm] &=&\pm \omega {\vec P}_\pm,\\
{} [H_\omega,{ L}_\pm] &=&\pm 2 \omega { L}_\pm.
\end{eqnarray}
The full set of commutation relations for the oscillator generators are given in Table~\ref{Comm2} of Appendix~\ref{sec:comm_relns}. Comparing Tables~\ref{Comm1} and~\ref{Comm2}, we see that the generators $\left(L_{+},{\vec P}_{+},H_{\omega},{\vec P}_{-},L_{-}\right)$ in oscillator space act correspondingly like $\left(H,{\vec P},D,{\vec K},C\right)$.   It follows from this table, as well as Eqs.~(\ref{eq:goo1})-(\ref{eq:goo2}) evaluated at $y=0$ that the state $|{\cal O}\rangle$ is a state of lowest weight, i.e.,
\begin{equation}
\vec{P}_{-}\left|\mathcal{O}\right\rangle =L_{-}\left|\mathcal{O}\right\rangle =0.
\end{equation}
The remaining states in the spectrum are obtained by raising with ${\vec P}_+$ and $L_+$.    These states, which are of the form
\begin{equation}
|\psi\rangle = {{ P}_+}{}_{i}\cdots L_+\cdots {{ P}_+}{}_{j}\cdots L_+\cdots|{\cal O}\rangle
\end{equation}
correspond to the descendants of the primary operator ${\cal O}(x)$ in the Galilean frame.   Thus the spectrum of $H_\omega$ states gets organized into towers of states evenly spaced in energy, each lying above a lowest weight state $|{\cal O}\rangle$.    When constructing descendants of $\left|\mathcal{O}\right\rangle $,
it is sometimes useful to work with generators
\begin{equation}
\label{eq:Q}
Q_{\pm}\equiv L_{\pm}-\frac{\vec{P}_{\pm}^{2}}{2N}.
\end{equation}
rather than $L_{\pm}$ defined above.   The nice feature of $Q_{\pm}$ is that they commute with both $\vec{P}_{+}$ and $\vec{P}_{-}$. A consequence is that descendants constructed by raising with different powers of $Q_{+}$ are orthogonal. For instance, the level two (i.e. two energy units above $|{\cal O}\rangle$) states $\left|Q\mathcal{O}\right\rangle \equiv Q_{+}\left|\mathcal{O}\right\rangle $ and ${{P}_{+}}{}_i {{P}_{+}}{}_j|{\cal O}\rangle$ are orthogonal.

While the Schrodinger algebra guarantees that there is an energy eigenstate for every primary ${\cal O}$, the logical possibility remains that there are states in the spectrum of $H_\omega$ that are not of the form Eq.~(\ref{eq:map}).   To close the loophole, we note that given an $N_{{\cal O}^\dagger}$-particle eigenstate $|{\cal O}\rangle$ of $H_\omega$ with eigenvalue $\omega\Delta_{\cal O}$ which is annihilated by ${ P}_{-i}$ and $L_-$, the operator
\begin{equation}
{\cal O}(0) \equiv |0\rangle\langle{\cal O}|e^{H/\omega},
\end{equation}
obeys $[C,{\cal O}(0)]=[{K_i},{\cal O}(0)]=0$ as well as $\left[N,\mathcal{O}(0)\right] = N_{\mathcal{O}}\mathcal{O}(0)$, $N_{\cal O}=-N_{{\cal O}^\dagger}<0$, and  $\left[D,\mathcal{O}(0)\right] = i\Delta_{\mathcal{O}}\mathcal{O}(0)$. So there is an inverse mapping from lowest weight states $|{\cal O}\rangle$ to primary operators.    
It follows that the spectrum of $H_\omega$ states is isomorphic to the space of primary operators and their descendants.

\subsection{Hilbert space interpretation of correlation functions}
\label{sec:hs}

Given that $|{\cal O}\rangle = \left[{\cal O}(-i/\omega,0)\right]^\dagger|0\rangle$ is a lowest weight state we may interpret the two-point function of primary operators as an inner product.   From Eq.~(\ref{eq:2pf}),
\begin{equation}
\langle{\cal O}_1|{\cal O}_2\rangle = \langle 0| {\cal O}_1(-i/\omega,0) \left[{\cal O}_2(-i/\omega,0)\right]^\dagger|0\rangle = G_2(-i/\omega,0;i/\omega,0)=c \delta_{\Delta_1,\Delta_2} \left(-{2i\over\omega}\right)^{-\Delta_1},
\end{equation}
so positivity of the norm $\langle{\cal O}|{\cal O}\rangle\geq 0$ fixes the normalization constant $c$ to be proportional to  $\left(-{2i\over\omega}\right)^{\Delta_1}$.  An immediate consequence of this inner product structure on operators is the unitarity bound on scaling dimensions, which follows from positivity of the state $\left|Q\mathcal{O}\right\rangle \equiv Q_+|\mathcal{O}\rangle.$  From the Schrodinger algebra, in the form given in Table~\ref{Comm3},
\begin{equation}
\left\langle Q\mathcal{O}|Q\mathcal{O}\right\rangle =\left(\Delta_{\mathcal{O}}-\frac{d}{2}\right)\left\langle \mathcal{O}|\mathcal{O}\right\rangle,
\end{equation}
implying that $\Delta_{\cal O}\geq d/2$ for any primary operator.    This result~\cite{Nishida:2010tm}, is the non-relativistic counterpart of the unitarity bound in relativistic CFTs.

Note that when $\Delta_{\mathcal{O}}=d/2$ saturates the unitarity bound, $\left|Q\mathcal{O}\right\rangle $ is a null state. One consequence is that the state created by $\tilde{\mathcal{O}}^{\dagger}(y)$ satisfies the harmonic oscillator Schrodinger equation.  This follows from the relation
\begin{equation}
\label{eq:Qc}
\left[Q_{\pm},\tilde{\mathcal{O}}(y)\right]=-\frac{1}{2\omega}e^{\mp2i\omega\tau}\left(i\partial_{\tau}-\frac{1}{2N_{\mathcal{O}}}\vec{\partial}_{y}^{2}+\frac{1}{2}N_{\mathcal{O}}\omega^{2}{\vec y}^{2}\pm\left(\Delta_{\mathcal{O}}-\frac{d}{2}\right)\omega\right)\tilde{\mathcal{O}}(y),
\end{equation}
which for $\Delta_{\mathcal{O}}=d/2$ implies that
\begin{equation}
\left(i\partial_{\tau}-\frac{1}{2N_{\mathcal{O}^{\dagger}}}\vec{\partial}_{y}^{2}+\frac{1}{2}N_{\mathcal{O}^{\dagger}}\omega^{2}y^{2}\right)\tilde{\mathcal{O}}^{\dagger}(\tau,\vec{y})\left|0\right\rangle =0.
\end{equation}
Even away from the unitarity bound $\Delta_{\mathcal{O}}=d/2$, there is a sense in which the state $|{\cal O}\rangle$ corresponds to a harmonic oscillator wavefunction.   Indeed, the two-point function in the oscillator frame, from Eqs.~(\ref{eq:2pf}), (\ref{eq:oxtoy}), is given by
\begin{equation}
\label{eq:o2pt}
\langle\cO\left(y_1\right)\cO^{\dagger}({y}_{2})\rangle=c\left(\sin{\omega\tau_{12}}\right)^{-\Delta_{\cO}}\exp{\left\{ -{i\over 2} N_{\cO}\omega\left[\left({\vec y}_{1}^{2}+{\vec y}_{2}^{2}\right)\cot{\omega\tau_{12}}-2\vec{y_{1}}\cdot\vec{y}_{2}\csc{\omega\tau_{12}}\right]\right\}},
\end{equation}
and therefore $\psi_{\cal O}(y)= \langle 0|{\tilde{\cal O}}(y)|{\cal O}\rangle$, 
\begin{eqnarray}
\nonumber
\psi_{\cal O}(y) = \langle 0|{\tilde{\cal O}}(y)\left[{\cal O}(-i/\omega,0)\right]^\dagger|0\rangle  &=& c\lim_{\tau_0\rightarrow -i\infty}\left[{\sin{\omega(\tau+\tau_0)}\over\cos\omega\tau_0}\right]^{-{\Delta_{\cal{O}}}} \exp\left\{-{1\over 2} N_{{\cO}^\dagger}\omega {\vec y}^2\right\}\\
&\propto& e^{-i\omega\Delta_{\cal O}\tau}\exp\left\{-{1\over 2} N_{{\cO}^\dagger}\omega {\vec y}^2\right\},
\end{eqnarray}
can be interpreted as the wavefunction for the center-of-mass position ${\vec y}$ of the many-body bound state $|{\cal O}\rangle$.   Evidently, the excited states generated by raising with $Q_+$ and ${\vec P}_+$ will have wavefunctions corresponding to the action of the differential operators on the RHS of Eqs.~(\ref{eq:goo1.5}), (\ref{eq:Qc}).   Given that $\psi_{\cal O}(y)$ is a harmonic oscillator ground state with energy $\Delta_{\cO}$, we have from Eq.~(\ref{eq:Qc}) 
\begin{equation}
\psi_{Q_+^n{\cO}}(y)= \langle 0|{\tilde{\cal O}}(y)Q_+^n|{\cal O}\rangle = \left(\Delta_{\cO} -d/2\right)_n e^{-2in\omega\tau} \psi_{\cal O}(y) \propto e^{-i\omega\left(\Delta_{\cal O}+2n\right)\tau} \exp\left\{-{1\over 2} N_{{\cO}^\dagger}\omega {\vec y}^2\right\}
\end{equation}
are also Gaussian.   On the other hand, from Eq.~(\ref{eq:goo1.5}),  the operators ${\vec P}_{\pm}$ act like usual harmonic oscillator raising and lowering operators ${\vec a}$, ${\vec a}^\dagger$ in quantum mechanics, and generate a tower of excited state wavefunctions above each state $Q_+^n|{\cO}\rangle$ with energy gap $\Delta E= 2\omega$.   Thus the operators ${\vec P}_\pm$ create and destroy the translational modes of $|{\cal O}\rangle$ while the operators $Q_+$ can be interpreted as exciting the internal degrees of freedom of the bound state.  Indeed, when $|{\cal O}\rangle$ saturates the unitary bound, these internal modes decouple, and the eigenstate spectrum coincides with that of a $d$-dimensional point particle in a harmonic potential.

The three-point function also has a natural interpretation in the Hilbert space description of the NRCFT.    From Eq.~(\ref{3pf}), we can relate the matrix elements $\langle {\cal O}_1|{\tilde{\cO}(y)}|{\cal O}_2\rangle$ to the correlator 
\begin{eqnarray}
\nonumber
\langle {\cal O}_1|{\tilde{\cO}(y)}|{\cal O}_2\rangle &=&  \langle 0|{\cal O}_1(-i/\omega,0) {\tilde{\cO}(y)} \left[{\cal O}_2(-i/\omega,0)\right]^\dagger|0\rangle\\
& =& 
\left[\cos\omega\tau\right]^{-{\Delta_{\cal{O}}}} \exp{\left[{i\over 2} N_{\cal O}\omega{\vec y}^2\tan\omega\tau\right]} G_3(-i/\omega,0;x;i/\omega,0),
\end{eqnarray}
or, in terms of the function $F(v)$ appearing on the RHS of Eq.~(\ref{3pf}),
\begin{equation}
\label{eq:ME}
\langle {\cal O}_1|{\tilde{\cO}(y)}|{\cal O}_2\rangle = 2^{\Delta_{\cO}} (i\omega)^{\Delta/2}e^{-i\omega(\Delta_2-\Delta_1)\tau} e^{-{1\over 2} N_{{\cO}^\dagger} \omega {\vec y}^2} F(v=i\omega {\vec y}^2).
\end{equation}
So we may interpret $F(v=i\omega {\vec y}^2)$ as a form factor for the local operator ${\tilde{\cO}(y)}$ between initial and final states $|{\cal O}_{1,2}\rangle$.   From Eqs.~(\ref{eq:goo1})-(\ref{eq:goo2}), the matrix elements of ${\tilde{\cO}(y)}$ between descendant states of $|{\cal O}_{1,2}\rangle$ are given by differential operators acting on the RHS of Eq.~(\ref{eq:ME}) and are therefore fixed in terms of derivatives of $F(v)$.    In the particular case where either ${\cO}_1$ or ${\cO}_2$ saturates the unitarity bound, it follows from Eq.~(\ref{eq:Qc}) that the function $F(v)$ appearing in the three-point function is fixed up to normalization.    As a consequence, from our results in sec.~\ref{sec:constraints}, the contribution of ${\cal O}(x)$ and its descendants to the ${\cal O}_1\times {\cal O}_2$ OPE is completely determined in this case.    These last two points were also noted in~\cite{Golkar:2014mwa}.

Higher-point correlation functions are related to the matrix elements
\begin{equation}
\langle {\cal O}_1|{\tilde\phi}_1(y_1)\cdots {\tilde\phi}_n(y_{n}) |{\cal O}_2\rangle.
\end{equation}
In the asymptotic limits $y_{ij}=y_i-y_j\rightarrow 0$ these functions can be expanded, via successive application of the OPE, as an infinite linear combination of matrix elements $\langle{\cal O}_1|{\tilde{\cO}}|{\cal O}_2\rangle$.    If the OPE expansion has a finite radius of convergence (rather than just being asymptotic, as in generic field theories) this observation implies that any matrix element in the Hilbert space of many-particle bound states is determined by the energy spectrum and by the set $\langle{\cal O}_1|{\tilde{\cO}}|{\cal O}_2\rangle$, at least locally.   Analogously, in the Galilean frame, OPE convergence would imply that the Green's functions in the charged sector are fixed in terms of the three-point functions and the operator spectrum.    The question then arises as to whether OPE convergence is an automatic consequence of Schrodinger invariance, and if so, how useful it is as an approximation scheme at finite $y_{ij}$.   We address these issues in sec.~\ref{sec:conv}.

\section{Bounds on OPE convergence}

\label{sec:conv}

By OPE convergence, we mean that between any two finite norm states $|\psi\rangle$, $|{\chi\rangle}$, the expansion
\begin{equation}
\label{eq:12OPE}
\lim_{y_1\rightarrow y_2} \langle\psi|{\tilde\phi}_1(y_1){\tilde\phi}_2(y_2)|\chi\rangle = \sum_{\alpha} c_\alpha(y_1,y_2)  \langle\psi| {\tilde\cO}_\alpha(y_1)|\chi\rangle
\end{equation}
has finite radius of convergence, determined by the coordinates $y_{1,2}$ and by the states $|\psi\rangle$, $|{\chi\rangle}$.    In this section we will argue that, under certain conditions, the OPE is a convergent expansion in NRCFTs, and that the rate of convergence is exponentially fast in operator dimensions, in a sense that we will explain below.   The arguments are very similar to the proof of convergence in four-dimensional CFTs in~\cite{Mack:1976pa} and more recently in~\cite{Pappadopulo:2012jk}.   In particular, as in the relativistic case, the completeness of the orthonormal basis of operators plays a crucial role.

Formally, OPE convergence in NRCFTs follows under the assumption that the set of states obtained by applying a string of local operators to the vacuum,
\begin{equation}
\label{eq:string}
{\tilde\phi}_1(y_1)\cdots {\tilde\phi}_n(y_n)|0\rangle,
\end{equation}
with non-zero net $U(1)_N$ charge, are normalizable states in the Hilbert space.    A particular such state is $|{\tilde\phi_1}\tilde{\phi_2},\chi\rangle ={\tilde\phi}_1(y_1){\tilde\phi}_2(y_2)|\chi\rangle$, since the finite norm state $|\chi\rangle$ itself can be expressed as a linear combination of imaginary time primary operators acting on the vacuum.    In the case of a product of operators whose charges do not add to zero, the expansion of  $|{\tilde\phi_1}\tilde{\phi_2},\chi\rangle$ into a basis of $H_\omega$ eigenstates can be carried out successively, by first performing the OPE as in Eq.~(\ref{eq:12OPE}) and then expanding each finite norm state ${\cal O}_\alpha(y_1)|{\chi}\rangle$ into energy eigenstates.  From the Hilbert space axioms, if  $|{\tilde\phi_1}\tilde{\phi_2},\chi\rangle$ has finite norm, it follows that its expansion into a complete set of orthonormal states converges to $|{\tilde\phi_1}\tilde{\phi_2},\chi\rangle$, see e.g.~\cite{ReedSimon}.  Because the eigenbasis expansion of ${\cal O}_\alpha(y_1)|{\chi}\rangle$ is  convergent, this chain of arguments then explains why the OPE is itself convergent.

The crucial step is then justifying finiteness of the norm of states of the form Eq.~(\ref{eq:string}).   In the case of CFTs, this was done using the principles of axiomatic field theory in refs.~\cite{Luscher:1975js,Mack:1976pa}.  A less rigorous approach, based on the path integral between radial quantization time slices was given in the textbook~\cite{Polchinski:1998rq} for the case of two dimensions, and generalized to higher dimensions in ref.~\cite{Pappadopulo:2012jk}.   
Briefly, from the path integral point of view, the existence (i.e. convergence) of the OPE follows from the fact that under radial quantization, the spatial slice on the cylinder corresponding to the infinite past gets mapped onto a single point, the origin of flat space. It is then plausible that specifying the infinite past state on the cylinder is equivalent to inserting an operator at the origin of flat space inside the path integral.  As we saw in Section \ref{sec:soc}, the NRCFT state-operator map is realized with a coordinate transformation that maps the infinite past in the Euclidean plane to a single point in the oscillator coordinates. Thus, the same path integral arguments used to justify the existence (i.e. convergence) of the OPE in CFTs should carry over to the NRCFT case as well.

\subsection{Estimating the convergence rate}
\label{sec:ConvRate}

In the relativistic case, ref.~\cite{Pappadopulo:2012jk} showed that the OPE converges exponentially fast, meaning that the error introduced by imposing a cutoff $E_{\star}$ on operator dimensions decays exponentially with $E_{\star}$.    In this section, we provide analogous bounds for NRCFTs.  We begin by noting that the states $|\psi\rangle$, $|{\chi\rangle}$ in Eq.~(\ref{eq:12OPE}) can be expanded in a basis of orthonormal $H_\omega$ eigenstates $\{|{\cal O}_\alpha\rangle\}$.   Since these are convergent expansions, to estimate the OPE convergence rate it is therefore sufficient to consider matrix elements of operators between these basis states, of the form
\begin{equation}
\label{eq:cme}
\langle{\cal O}_\alpha| {\tilde\phi}_1(y_1){\tilde\phi}_2(y_2)|{\cal O}_\beta\rangle.
\end{equation}
This matrix element can be interpreted as the inner product of the states $|1;\alpha\rangle={\tilde\phi}^\dagger_1(y_1)|{\cal O}_\alpha\rangle$ and $|2;\beta\rangle = {\tilde\phi}_2(y_2)|{\cal O}_\beta\rangle$, so by the Cauchy-Schwartz inequality, Eq.~(\ref{eq:cme}) is bounded by
\begin{equation}
|\langle{\cal O}_\alpha| {\tilde\phi}_1(y_1){\tilde\phi}_2(y_2)|{\cal O}_\beta\rangle|^2\leq |\langle1;\alpha|1;\alpha\rangle|^2   |\langle2;\beta|2;\beta\rangle|^2. 
\end{equation}
In Hilbert space, absolute convergence implies convergence, so we need to establish a bound on four-point functions 
\begin{equation}
 |\langle1;\alpha|1;\alpha\rangle|^2 = \langle 0|{\cal O}_\alpha(-i/\omega,0){\tilde\phi}(y) \left[{\tilde\phi}(y)\right]^\dagger \left[{\cal O}_\alpha(-i/\omega,0)\right]^\dagger|0\rangle.
\end{equation}

To simplify the chain of reasoning, we will consider the OPE for imaginary time $\tau$, so we work with a Euclidean time coordinate $\theta=i\omega\tau$, and assume $\theta>0$.  Then we must bound the matrix element
\begin{equation}
\label{eq:g4}
G_4(\alpha,{\tilde\phi};\theta,{\vec y})=|\langle1;\alpha|1;\alpha\rangle|^2 =\langle{\cO}_\alpha| {\tilde\phi}(\theta,{\vec y})  {\tilde\phi}^\dagger(-\theta,{\vec y})|{\cO}_\alpha\rangle.
\end{equation}
Note that, being the norm of a state in the NRCFT Hilbert space, this quantity is real and positive definite.

Following~\cite{Pappadopulo:2012jk}, it is useful to perform a spectral decomposition, first by expanding the state ${\tilde\phi}^\dagger(-\theta,{\vec y})|{\cO}_\alpha\rangle$ in the basis of energy eigenstates.    This expansion takes the general form,
\begin{equation}
\label{eq:eexp}
 {\tilde\phi}^\dagger(-\theta,{\vec y})|{\cO}_\alpha\rangle= \sum_{\cal O}\sum_{r,s,q} e^{-\theta\left(\Delta_{\cal O}-\Delta_{\cO_\alpha} + r + 2s + 2q\right)} f^{\cal O}_{rsq}({\vec y})|{\cal O}_{r,s,q}\rangle,
\end{equation}
where the outer sum runs over all primary operators, and the (normalized) states $|{\cal O}_{r,s,q}\rangle$, with energy $\omega(\Delta_{\cal O}+r+2s+2q)$ are proportional to $({\hat y}\cdot {\vec P}_+)^r \left({\vec P}_+^2\right)^s (Q_+)^q|{\cO}\rangle$.    In this equation, the exponential dependence on time is trivially fixed by Eq.~(\ref{eq:goo1}), while the spatial dependence, encoded in the coefficient functions $f^{\cO}_{rsq}({\vec y})$, depends on the properties of ${\tilde\phi}(y)$ and $|{\cO}_\alpha\rangle$.   It is useful to organize the sum over states on the RHS by energy level $n=r+s+2q$.   Then the above equation can be written as  
\begin{equation}
 {\tilde\phi}^\dagger(-\theta,{\vec y})|{\cO}_\alpha\rangle =  e^{-\theta\left(\Delta_{\cal O}-\Delta_{\cO_\alpha}\right)} \sum_{\cO}\sum_{n=0}^\infty e^{-n\theta} \sum_{\{\sigma\}} f^{\cal O}_{n;\{\sigma\}}({\vec y}) |{\cal O};n,\sigma\rangle,
\end{equation}
where $\{|{\cal O};n,\sigma\rangle\}$ is an orthonormal basis, indexed by level $n$ and a pair of integers $\{\sigma\}$ whose precise relation to the original set $r,s,q$ is not needed in what follows.  Inserting this into the four-point function then gives the required spectral decomposition
\begin{equation}
G_4(\alpha,{\tilde\phi};\theta,{\vec y}) = e^{-2 \theta\left(\Delta_{\cal O}-\Delta_{\cO_\alpha}\right)}\cdot\int_0^\infty dE e^{-2\theta E} \rho(E,{\vec y}),
\end{equation}
with spectral function 
\begin{equation}
\rho(E,{\vec y})= \sum_{{\cO},n,\{\sigma\}} \delta(E-n) \left|f^{\cal O}_{n;\{\sigma\}}({\vec y})\right|^2.
\end{equation}

As in the case of relativistic CFTs, it is possible to estimate the convergence rate of the OPE even without detailed knowledge of the spectral density $\rho(E,{\vec y})$, in particular the functions $f^{\cal O}_{n;\{\sigma\}}({\vec y})$.   To this end, we use the OPE $\tilde{\phi}(\theta,{\vec y})\times \tilde{\phi}^\dagger(-\theta,{\vec y})$ in the limit\footnote{ Note that even though this OPE does not involve an expansion in terms of charged primary operators, it follows from the Cauchy-Schwartz inequality that it also has a finite radius of convergence.   Indeed, arbitrary (Euclidean time) matrix elements $\langle\psi|{\tilde\phi}(y_1){\tilde\phi}^\dagger(y_2)|\chi\rangle$ are bounded by 
\begin{equation}
 |\langle\psi|{\tilde\phi}(y_1){\tilde\phi}^\dagger(y_2)|\chi\rangle|^2 \leq |\langle\psi|{\tilde\phi}(\theta_1,{\vec y}_1){\tilde\phi}^\dagger(-\theta_1,{\vec y}_1)|\psi\rangle|^2  |\langle\chi|{\tilde\phi}(\theta_2,{\vec y}_2){\tilde\phi}^\dagger(-\theta_2,{\vec y}_2)|\chi\rangle|^2.
 \end{equation}
But the two matrix elements on the RHS are precisely the norm of states of the form Eq.~(\ref{eq:eexp}), and therefore finite.   Thus the limit $\theta\rightarrow 0^+$ of the LHS, given by inserting the OPE $\tilde{\phi}(\theta,{\vec y})\times \tilde{\phi}^\dagger(-\theta,{\vec y})$, is also finite.} $\theta\rightarrow 0^+$.  Even though the terms appearing in the $\tilde{\phi}(\theta,{\vec y})\times \tilde{\phi}^\dagger(-\theta,{\vec y})$ OPE cannot be decomposed into Schrodinger group representations, for $\theta<\theta_0$ sufficiently small, the leading (singular) term, proportional to the identity operator, gives a good approximation
\begin{equation}
\label{eq:zOPE}
\lim_{\theta\rightarrow 0^+} \tilde{\phi}(\theta,{\vec y})\times \tilde{\phi}^\dagger(-\theta,{\vec y})\sim \left[\langle 0|\tilde{\phi}(\theta,{\vec y}) \tilde{\phi}^\dagger(-\theta,{\vec y})|0\rangle\right]\mathbb{I} \sim {c\over(-i\theta)^{\Delta_{\phi}}} \cdot \mathbb{I}.
\end{equation}
Note that the RHS is positive definite, given that $c$ is proportional to $(i\omega)^{-\Delta_{\phi}}$ (recall the discussion in sec.~\ref{sec:hs}).   Inserting this OPE into the four-point correlator $G_4(\alpha,{\tilde\phi};\theta,{\vec y})$ allows one to obtain a conservative bound on the remainder term
\begin{equation}
R\left(\theta,{\vec y};E_{\star}\right)=\int_{E_{\star}}^{\infty}dE\, e^{-2 \theta E}\, \rho(E,{\vec y}),
\end{equation}
with $E_{\star}\rightarrow\infty$ and $(\theta,{\vec y})$ fixed.    For a sufficiently small $\theta'<\theta$, we have
\begin{equation}
R\left(\theta,{\vec y};E_{\star}\right)\leq e^{-2 (\theta-\theta')E_{\star}} R(\theta',{\vec y};E_{\star}) \leq e^{-2 (\theta-\theta')E_{\star}} R(\theta',{\vec y};E_{\star}=0),
\end{equation}
where the last inequality follows from positivity of the spectral function.   If $\theta'<\theta_0$, the use of Eq.~(\ref{eq:zOPE}) implies that
\begin{equation}
R\left(\theta,{\vec y};E_{\star}\right)\leq {|c|\over (\theta')^{\Delta_{\phi}}}  e^{-2 (\theta-\theta')E_{\star}}. 
\end{equation}
The bound is optimized by choosing $\theta' = \Delta_\phi/2E_{\star}$, so the estimate for the remainder function is
\begin{equation}
\label{eq:bd}
R\left(\theta,{\vec y};E_{\star}\right)\leq \left(|c| \Delta_\phi^{-\Delta_\phi} e^{\Delta_\phi} \right) E_{\star}^{\Delta_\phi} e^{-2\theta E_{\star}},
\end{equation}
valid for $E_{\star}>{\Delta_\phi/\theta_0}$, $\theta_0<\theta$.

The bound in Eq.~(\ref{eq:bd}) implies that when the OPE is inserted inside a correlation function or matrix element, the error in dropping operators of dimension $\Delta>E_{\star}$ is exponentially small in $E_{\star}$, with the parameter $\theta$ that controls the radius of convergence determined by the location of the nearest operator insertion (see the examples in the next section).   Note that our bound is only conservative.    The optimal estimate can be obtained by using the mathematical machinery of the (so called) Hardy-Littlewood Tauberian theorem for Laplace transforms, see ref.~\cite{Korevaar:2004} (sections I.15 and VII.2).   This theorem gives the same result as above, but with the optimal constant $|c|/\Gamma\left(1+\Delta_{\phi}\right)$.  It also provides the next-to-leading term in the asymptotic behavior of the remainder, which is suppressed by a factor of $1/\ln E_{\star}$.

\subsection{Consistency checks}
\label{sec:3PF}

The arguments used in the previous section reduce the convergence properties of the OPE inserted in any $n$-point correlation function (or equivalently between finite norm states) to convergence inside four-point correlators such as the one in Eq.~(\ref{eq:g4}).  For example, a three-point function ${\tilde G}_3(y_1,y_2,y_3)=\langle  {\tilde\phi}_1(y_1)  \left[{\tilde\phi}_2(y_2)\right]^\dagger  \left[{\tilde\phi}_3(y_3)\right]^\dagger\rangle$ can be regarded as the inner product of the states $\left[{\tilde\phi}_1\right]^\dagger(y_1)|0\rangle$ and $ {\tilde\phi}_2(y_2)  {\tilde\phi}_3(y_3)|0\rangle$.   This inner product is bounded by 
\begin{equation}
|G_3(y_1,y_2,y_3)|^2 \leq |\langle 0|{\tilde\phi}_1(y_1)\left[{\tilde\phi}_1(y_1)\right]^\dagger|0\rangle|^2   |\langle 0|{\tilde\phi}_3(y_3){\tilde\phi}_2(y_2)\left[{\tilde\phi}_2(y_2)\right]^\dagger \left[{\tilde\phi}_3(y_3)\right]^\dagger|0\rangle|^2. 
\end{equation}
The first factor on the RHS is the two-point function, given in Eq.~(\ref{eq:o2pt}), thus the convergence rate of the ${\tilde\phi}^\dagger_2\times {\tilde\phi}^\dagger_3$ OPE inside ${\tilde G}_3(y_1,y_2,y_3)$ is controlled by a four-point function.   Using translation invariance in Euclidean time, we may set $\theta_3=-\infty$, so that the three-point function is bounded by the same type of matrix element,
\begin{equation}
\langle{\tilde\phi}_3| {\tilde\phi}_2(\theta_2,{\vec y}_2) {\tilde\phi}^\dagger_2(-\theta_2,{\vec y}_2)|{\tilde\phi}_3\rangle,
\end{equation}
analyzed above.   It follows that the OPE converges exponentially inside the three-point function.  Note, though, that because two-point functions are diagonal, inserting the OPE into a three-point function only yields contributions from a single primary and its descendants.    Thus convergence inside the three-point correlator only refers to how quickly descendant contributions fall off within a fixed conformal multiplet.   

Because their convergence hinges on the convergence properties of four-point correlators, we may use the three-point functions in known NRCFTs as a consistency check of our arguments in the previous sections.  In particular, if these three-point correlators fail to exhibit exponential convergence then so do the four-point function, or any other correlator for that matter.    To this end, we now analyze the convergence of the OPE inside specific three-point functions, in the context of several explicit model NRCFTs.    Besides the theory of a free boson, we consider the strongly interacting NRCFT of bosons tuned to $S$-wave unitarity in $d=3$ space dimensions, as well as a toy holographic model~\cite{Volovich:2009yh}, with no known many-body quantum field theory realization.   For all these examples, we work in Galilean coordinates, and cutoff the terms in the OPE at some large dimension $E_{\star}$.   In all these examples, we find, using conservative estimates, that the discarded terms in the OPE, i.e. the tail $T_{E_{\star}}$ of the OPE sum inside $G_3(x_1,x_2,x_3)$, universally decay at least as fast as 
\begin{equation}
\left|T_{E_{*}}\right|\lesssim E_{*}^{-\frac{1}{2}}e^{-E_{*}\beta},\qquad\qquad E_{*}\gg1,\label{3PF Rate}
\end{equation}
\noindent where $\beta$ is a \emph{universal} (i.e. model independent) function of $t_{21}/t_{31}$, as well as the ratio ${\vec x}_{21}^{2}/t_{21}$ which we hold fixed in the OPE limit $x_1\rightarrow x_2$. The bound exhibits the expected exponential suppression in $E_{\star}$.

\subsubsection{Free field theory}
\label{subsec:3PFFT}

First we consider the free boson of Eq.~(\ref{eq:free}).    We denote the boson charge by $N=-1$.    As an example, we analyze the asymptotic behavior of the $\phi\times\phi$ OPE inserted into the three-point function with the primary operator $(\phi^2)^\dagger$, in the limit $x_2\rightarrow 0$ with ${\vx}^2_2/t_2$ and $x_3$ fixed.   In the Galilean frame, the correlator is proportional to two insertions of the $\langle \phi\phi^\dagger\rangle$ propagator
\begin{equation}
\left\langle [\phi^2]^{\dagger}(x_3)\phi(x_2)\phi(0)\right\rangle \propto t_{3}^{-\Delta}\text{exp}\left[N\frac{{\vx}_{3}^{2}}{2t_{3}}\right]\times t_{23}^{-\Delta}\text{exp}\left[-N\frac{{\vx}_{23}^{2}}{2t_{23}}\right],\label{Free3PF}
\end{equation}
with $\Delta=d/2$ (note that we are working in Euclidean time). 

The tail $T_{E_{\star}}$ is defined as the contribution to the Green's function from the OPE, but involving operators of dimension larger than $\Delta>E_{\star}$.   To compute this quantity it is sufficient to Taylor expand the second propagator in Eq.~(\ref{Free3PF}) in the limit $x_2\rightarrow 0$, keeping terms with more than $\sim E_{\star}$ derivatives (keeping in mind that a time derivative has scaling dimension two while a space derivative has scaling dimension one).   Some useful Taylor
expansion formulas are given in Appendix~\ref{app:Taylor}.  The result, dropping the prefactor $t_{3}^{-\Delta}\text{exp}\left[N\frac{{\vx}_{3}^{2}}{2t_{3}}\right]$,
is  
\begin{eqnarray}
&&T_{E_{\star}}=\sum_{E=E_{\star}}^{\infty}\left(\frac{t_{2}}{t_{3}}\right)^{\frac{E}{2}}\sum_{p=0}^{\left\lfloor E/2\right\rfloor }\sum_{n=0}^{\left\lfloor E/2-p\right\rfloor }\sum_{r=0}^{p}\frac{\left(-1\right)^{n-r}N^{E-p-n-r}}{2^{n+p-r}p!n!\left(E-2p-2n\right)!}\begin{pmatrix}p\\
r
\end{pmatrix}\frac{\Gamma(\Delta+E-p-n)}{\Gamma(\Delta+E-p-n-r)} \times \nonumber \\
&&\hspace{65mm} \left(-\frac{{\vx}_{2}^{2}}{t_{2}}\right)^{n}\left(-\frac{{\vx}_{3}^{2}}{t_{3}}\right)^{p-r}\left(\frac{(\vec{x}_{2}\cdot \vec{x}_{3})^{2}}{t_{2}t_{3}}\right)^{\frac{E}{2}-p-n},\label{FreeTail}
\end{eqnarray}
\noindent where $\left\lfloor k\right\rfloor $ is the greatest integer
less than or equal to $k$. In the limit $x_2\rightarrow 0$ with ${\vx_{2}}^{2}/t_{2}$ fixed, one has
\begin{equation}
\left|T_{E_{\star}}\right|\leq\sum_{E=E_{\star}}^{\infty}\left|\frac{t_{2}}{t_{3}}\right|^{\frac{E}{2}}\sum_{p=0}^{\left\lfloor E/2\right\rfloor }\sum_{n=0}^{\left\lfloor E/2-p\right\rfloor }\sum_{r=0}^{p}\frac{\left|Nz\right|^{E-p-n-r}}{2^{n+p-r}p!n!\left(E-2p-2n\right)!}\begin{pmatrix}p\\
r
\end{pmatrix} \frac{\Gamma(\Delta+E-p-n)}{\Gamma(\Delta+E-p-n-r)} \label{FreeBound1},
\end{equation}
\noindent where
\begin{equation}
z\equiv\text{max}\left\{ \left|\frac{{\vx_{2}}^{2}}{t_{2}}\right|,\left|\frac{{\vx}_{3}^{2}}{t_{3}}\right|\right\} .\label{Freez}
\end{equation}
\indent At this stage, we make two conservative bounds. First, the ratio of
Gamma functions appearing above can be bounded by
\begin{equation}
\frac{\Gamma(\Delta+E-p-n)}{\Gamma(\Delta+E-p-n-r)}\leq\frac{\Gamma(\Delta+E+1)}{\Gamma(\Delta+E/2+1)}.\label{GammaBound}
\end{equation}
\noindent Second, the factor $\left[\left(E-2p-2n\right)!\right]^{-1}$
can be rewritten using $(2k)!=2^{k}k!(2k-1)!!$ and bounded by dropping
the double factorial altogether. With these bounds in place, we can
do the sums over $r$, $n$, and $p$ to get 
\begin{equation}
\left|T_{E_{\star}}\right|\leq\sum_{E=E_{\star}}^{\infty}\left|\frac{t_{2}}{t_{3}}\left(1+\left|Nz\right|+\frac{1}{2}\left|Nz\right|^{2}\right)\right|^{\frac{E}{2}}\frac{\Gamma(\Delta+E+1)}{\Gamma(\Delta+E/2+1)\Gamma(E/2+1)}\label{FreeBound2}.
\end{equation}
\indent Finally, using Stirling's approximation, $k!\sim\sqrt{2\pi}k^{k+1/2}e^{-k}$
for large $k$, we get that $\left|T_{E_{\star}}\right|$ decays as
\begin{equation}
\left|T_{E_{*}}\right|\lesssim E_{*}^{-\frac{1}{2}}e^{-E_{*}\beta},\qquad\qquad E_{*}\gg1,\label{FreeBoundResult}
\end{equation}
\noindent where
\begin{equation}
\beta\equiv\frac{1}{2}\text{log}\left|\frac{1}{\left(4+4\left|Nz\right|+2\left|Nz\right|^{2}\right)}\frac{t_{3}}{t_{2}}\right|,\label{FreeBeta}
\end{equation}
\noindent and we have dropped an $E_{*}$-independent prefactor $2^{\Delta}\sqrt{\frac{2}{\pi}}\left[1-e^{-\beta}\right]^{-1}$
in Eq.(\ref{FreeBoundResult}).\\
\indent We reiterate that this is a conservative estimate for the convergence
rate and the radius of convergence, the latter given by $|t_{2}| < (4+4\left|Nz\right|+2\left|Nz\right|^{2})^{-1} |t_{3}|$. 
Note that the radius of convergence is controlled by $|t_{2}/t_{3}|$. 

\subsubsection{Bosons at unitarity}
\label{3PFUBG}

\begin{figure}
\centering
\includegraphics[trim=0in .5in 0in .25in,clip,scale=0.54]{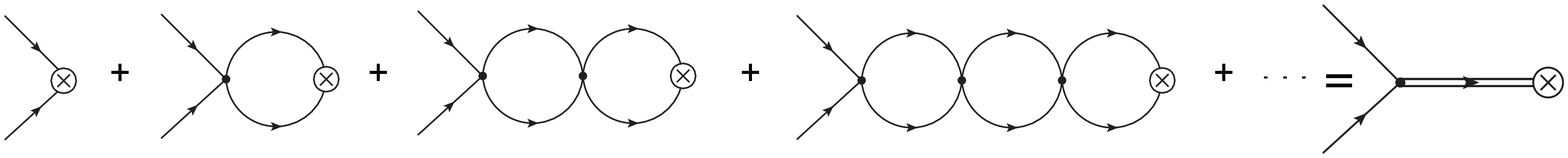}
\caption{The three-point function involving two $\phi$ insertions and one insertion of $\Phi=(\phi^2)^\dagger$ is given by the sum of loop diagrams shown in the figure. The sum of diagrams is equivalent to the exchange of a charge $-2$ field of dimension $\Delta=2$.}
\label{fig:bubbles}
\end{figure}

By bosons at unitarity, we mean a non-relativistic field theory of bosons $\phi$ in $d=3$ spatial dimensions with Lagrangian
\begin{equation}
\label{UBG L}
{\cal L} = \phi^{\dagger}\left(i\partial_{t}+\frac{\nabla^{2}}{2}\right)\phi - {\lambda\over 4} (\phi^\dagger\phi)^2,
\end{equation}
with the coupling constant $\lambda$ tuned such that the two-body $S$-wave scattering amplitude saturates the unitarity bound.    In this limit, with $S$-wave scattering length $|a|\rightarrow\infty$, the theory becomes Schrodinger invariant\footnote{Strictly speaking, the Lagrangian in Eq.~(\ref{UBG L}) is only Schrodinger invariant in the two-body sector. In sectors with more than two particles, scaling symmetry is broken to a discrete group as a consequence of the Efimov effect~\cite{Efimov:1970zz,*Efimov:1971zz}. Alternatively, in $d=3$, one could consider spin-1/2 fermions with an $S$-wave contact interaction tuned to unitarity.   In that case, the three-point correlator would take a form identical to Eq.~(\ref{UBG 3PF}) (up to spin labels and numerical prefactors), while three-body Efimov limit cycles would be forbidden by fermion statistics.}~\cite{Mehen:1999nd}.   The value of $\lambda$ at this fixed point depends on the ultraviolet regulator, for example in dimensional regularization, the unitary limit is $\lambda\rightarrow\infty$.  It is straightforward to show, by summing bubble diagram chains such as the one in fig.~\ref{fig:bubbles}, that at the strongly interacting fixed point, $\phi$ saturates the unitarity bound $\Delta_\phi=3/2$, the operator $\phi^2$ is a primary of dimension $\Delta_{\phi^2}=2$, and the three-point correlator of $(\phi^2)^\dagger$, $\phi$, and $\phi$ is~\cite{Fuertes:2009ex}
\begin{equation}
\left\langle [\phi^2]^{\dagger}(x_3)\phi(x_2)\phi(0)\right\rangle \propto t_{2}^{-\frac{1}{2}}t_{3}^{-1}\text{exp}\left[\frac{{\vx}_{3}^{2}}{2t_{3}}\right]\times t_{23}^{-1}\text{exp}\left[-\frac{{\vx}_{23}^{2}}{2t_{23}}\right]\times v^{-\frac{1}{2}}\gamma\left(\frac{1}{2},v\right),\label{UBG 3PF}
\end{equation}
\noindent where $t_3<t_2$, and $v\equiv v_{123}$ was defined in Eq.~(\ref{vij}), and $\gamma(s,x)$ is the lower incomplete gamma function,
so that 
\begin{equation}
v^{-\frac{1}{2}}\gamma\left(\frac{1}{2},v\right)=\int_{0}^{1}dw\: w^{-1/2}\text{exp}\left[-vw\right].\label{incomplete gamma}
\end{equation}
\indent Eq.~(\ref{UBG 3PF}) is a product of three factors. The first has trivial
$x_2$-dependence. The second factor is a propagator, which
we have already determined above to exhibit exponential convergence.
Thus, it remains to examine the third factor. This can be done by
starting with Eq.~(\ref{incomplete gamma}) and using the Taylor expansion
formulas in Appendix~\ref{app:Taylor} to expand $\text{exp}\left[-vw\right]$ in
$x_2$ about $x_1=0$. The only subtlety is that in this
limit $v\rightarrow -{\vx_{2}}^{2}/2t_{2}$, which we are keeping fixed,
so it is useful to pull out a factor of $\text{exp}\left[-w{\vx}_{2}^{2}/2t_{2}\right]$
before Taylor expanding.  One finds $T_{E_{*}}$
to be given again precisely by the form in Eq.~(\ref{FreeTail}), except for the following extra integral appearing in the
summand:
\begin{equation}
\int_0^1 dw\: w^{-1/2}w^{E-p-n-r}\text{exp}\left[w{\vx}_{2}^{2}/2t_{2}\right].
\end{equation}
\noindent Since in the sum $p+n+r\leq E$, the absolute value of this
integral is bounded (by $2$). Up to a constant factor, then, one
recovers Eq.~(\ref{FreeBound1}), which demonstrates exponential convergence
for the $v^{-1/2}\gamma(\frac{1}{2},v)$ factor and thus for the correlator
in Eq.~(\ref{UBG 3PF}) as a whole.

\subsubsection{Holographic model}
\label{subsec:3PFHolo}

As a final example, we consider NRCFTs arising as AdS/CFT duals of gravity theories that propagate in a Schrodinger invariant background spacetime~\cite{Balasubramanian:2008dm,Son:2008ye}.   Even though (to our knowledge) the NRCFTs that arise in this way have no known Lagrangian description, these theories have correlators that by the standard AdS/CFT dictionary are Schrodinger invariant, so must obey the constraints discussed above.   For example, the three-point function of primary scalars in such models, computed in~\cite{Volovich:2009yh} assuming generic bulk cubic scalar interactions, is given by     
\begin{equation}
\left\langle \mathcal{O}^{\dagger}(x_3)\mathcal{O}_{2}(x_2)\mathcal{O}_{1}(x_1)\right\rangle \propto t_{12}^{-\Delta_{12,3}/2}t_{13}^{-\Delta_{13,2}/2}\text{exp}\left[N_{1}\frac{x_{13}^{2}}{2t_{13}}\right]\times t_{23}^{-\Delta_{23,1}/2}\text{exp}\left[N_{2}\frac{x_{23}^{2}}{2t_{23}}\right]\times I(v),\label{Holo3PF}
\end{equation}
\noindent where $v=v_{123}$ again, and $I(v)$ is
the following integral,%
\footnote{In~\cite{Volovich:2009yh}, $I(v)$ is initially defined as a particular double contour
integral, with the expression in Eq.(\ref{I(v)}) obtained after some
evaluation. In particular, Eq.(\ref{I(v)}) assumes $t_{12},t_{23}>0$.%
}
\begin{equation}
I\left(v\right)\equiv C\int_{0}^{1}dz\: z^{\frac{\Delta_{12,3}}{2}-1}\left(1-z\right)^{\frac{\Delta_{13,2}}{2}-1}\left(1+\frac{N_{1}}{N_{2}}z\right)^{\frac{\Delta_{23,1}}{2}-1}\text{exp}\left[iN_{1}zv\left(1+i\epsilon\right)\right],\qquad\left|\frac{N_{1}}{N_{2}}\right|<1\label{I(v)}
\end{equation}
\begin{equation}
C\equiv\frac{4\pi^{2}N_{1}^{\Delta_{1}-1}N_{2}^{\frac{\Delta_{23,1}}{2}-1}}{\Gamma\left(\frac{\Delta_{12,3}}{2}\right)\Gamma\left(\frac{\Delta_{13,2}}{2}\right)\Gamma\left(\frac{\Delta_{23,1}}{2}\right)}\text{exp}\left[-i\frac{\pi}{4}\sum_{i}\Delta_{i}\right].\label{C}
\end{equation}
\indent As in the unitary bosons example considered above, Eq.~(\ref{Holo3PF}) is a product of three factors, where the first
factor has trivial $x_2$ dependence and the second factor
is a propagator whose expansion we have already shown converges exponentially.
It remains to understand the convergence rate of the Taylor
expansion of $I(v)$. In particular, we need to expand the exponential
in Eq.(\ref{I(v)}) in $x_2\rightarrow x_1$. We have
already encountered such an expansion in the preceding example.
It follows that the tail $T_{E_{*}}$ of the Taylor series of $I(v)$
is again given by the form in Eq.(\ref{FreeTail}), except
for the following extra integral appearing in the summand:
\begin{equation}
\int_{0}^{1}dz\: z^{\frac{\Delta_{12,3}}{2}-1}\left(1-z\right)^{\frac{\Delta_{13,2}}{2}-1}\left(1+\frac{N_{1}}{N_{2}}z\right)^{\frac{\Delta_{23,1}}{2}-1}z^{E-p-n-r}\text{exp}\left[iN_{1}z\frac{x_{12}^{2}}{2t_{12}}\left(1+i\epsilon\right)\right],\qquad\left|\frac{N_{1}}{N_{2}}\right|<1
\end{equation}
\noindent As in the unitary bosons example, this integral is bounded
since $E-p-n-r\geq0$, and so up to a constant factor one recovers
Eq.~(\ref{FreeBound1}), which demonstrates exponential convergence
for $I(v)$ and thus for the correlator as a whole.

\section{Discussion}
\label{sec:conclusion}

Motivated by applications to atomic systems in the unitary regime, in this paper we have initiated a systematic analysis of the OPE in NRCFTs.    We have found that in such theories, the OPE has a structure analogous to what is found in their relativistic counterparts.   In particular, for charged operators, the OPE organizes itself into contributions from conformal multiplets, consisting of a primary operator and its tower of descendants.    In the charged sector, the OPE converges exponentially fast, meaning that the error term resulting from truncating the OPE at dimension $E_{\star}\gg 1$ is exponentially small in $E_{\star}$.     An important implication is that correlation functions involving charged operators can be well-approximated (with a quantifiable error) by inserting the OPE and keeping only the leading low-dimension operators.    Additionally, it follows from OPE convergence that $n$-point Green's functions are determined in terms of the two-point correlators, which are fixed by symmetry, and the three-point functions, which contain dynamical information in the form of a single function of the three-point conformal invariant.

Unfortunately, non-relativistic conformal invariance seems to have little to say about the OPE in the physically important case of operators with vanishing particle number, for example conserved currents and the stress tensor.     The reason is that the Hilbert space interpretation of operators in the oscillator frame, which played a crucial role in our analysis, breaks down in the sector of operators with $N_{\cal O}=0$, as explained in~\cite{Bekaert:2011qd}.   However, our results open the possibility of constraining the charged sector of NRCFTs, i.e. bound state spectra and matrix elements, by methods analogous to the conformal bootstrap in relativistic theories.

A possible way for implementing this NRCFT bootstrap could be to take the limit of vanishing spatial coordinates.   In that limit, the kinematics becomes the same as in $d=1$ CFTs (conformal quantum mechanics).    Then the four-point correlators are functions of the invariant $u=t_{12} t_{34}/t_{13} t_{24}$ and using crossing symmetry and the OPE, it is possible to obtain sum rules analogous to the usual bootstrap-type equations as in CFTs (the conformal bootstrap in $0+1$-dimensional CFT has been discussed in~\cite{Gaiotto:2013nva,Golden:2014oqa}).   One possible obstacle to carrying out this program is that in the NRCFT case, the OPE Wilson coefficients are derivatives of an unknown function $F(z={\vx}^2/t)$ appearing in the three-point function.   But the results of this paper imply that $F(z)$ is analytic at $z=0$, so exponential convergence of the descendant expansion allows us to approximate $F(z\rightarrow 0)$ by a finite-degree polynomial in $z$.    The number of unknown degrees of freedom per primary operator is then effectively finite (to reasonable numerical accuracy), as in the CFT case.  

Such a scheme would probably be most useful in the sector of operators $N_{\cal O}\gg 1$, for which numerical integration of the $N_{\cal O}$-body Schrodinger equation is not possible.   Given that the number of degrees of freedom in the non-relativistic bootstrap is bound to be larger than for CFTs, it remains to be seen if the procedure sketched here is numerically feasible.   A natural setting for testing these ideas would be the two- and three-body sector of $d=3$, spin-1/2 fermions at unitarity, where much is already known, e.g.~\cite{Nishida:2007pj,PhysRevA.78.013614}, about the spectrum of scaling dimensions.

\begin{acknowledgments}

We thank Andrew Cohen, Emanuel Katz, Riccardo Rattazzi, and Sergey Sibiryakov for helpful discussions. This work was supported in part by DOE grant DE-FG02-92ER-40704 (W. G. and S. P.), and DOE grant DE-SC0010025 (Z. K.).

\end{acknowledgments}
\appendix

\section{NRCFT commutation relations}
\label{sec:comm_relns}
Tables~\ref{Comm1}-\ref{Comm3} below contain the commutation relations for the NRCFT generators in different bases. Not included are the number operator $N$, which commutes with everything, and the rotation generator $M_{ij}$, whose commutation relations with the other generators are determined as usual by their properties under rotation. Recall that
\begin{eqnarray}
H_{\omega} &\equiv& H+\omega^{2}C, \\
\vec{P}_{\pm} &\equiv& \frac{1}{\sqrt{2\omega}}\vec{P}\pm i\sqrt{\frac{\omega}{2}}\vec{K} \\
L_{\pm} &\equiv& \frac{1}{2}\left(\frac{1}{\omega}H-\omega C\pm iD\right) \\
Q_{\pm} &\equiv& L_{\pm}-\frac{\vec{P}_{\pm}^{2}}{2N}.
\end{eqnarray}
The advantage of the last basis is that $Q_\pm$ commute with $\vec{P}_\pm$. 
\begin{table}[h]
\caption{Commutation relations $[X,Y]$ in the $\left(H,\vec{P},D,\vec{K},C\right)$ basis.}
\begin{tabular}{|c|c|c|c|c|c|}
\hline 
\hspace{5mm}$X\backslash Y$\hspace{5mm} & \hspace{5mm}$P_{j}$\hspace{5mm} & \hspace{5mm}$K_{j}$\hspace{5mm} & \hspace{5mm}$H$\hspace{5mm} & \hspace{5mm}$C$\hspace{5mm} & \hspace{5mm}$D$\hspace{5mm} \tabularnewline
\hline 
$P_{i}$ & 0 & $-i\delta_{ij}N$ & 0 & $-iK_{i}$ & $-iP_{i}$\tabularnewline
\hline 
$K_{i}$ & $i\delta_{ij}N$ & 0 & $iP_{i}$ & 0 & $iK_{i}$\tabularnewline
\hline 
$H$ & 0 & $-iP_{j}$ & 0 & $-iD$ & $-2iH$\tabularnewline
\hline 
$C$ & $iK_{j}$ & 0 & $iD$ & 0 & $2iC$\tabularnewline
\hline 
$D$ & $iP_{j}$ & $-iK_{j}$ & $2iH$ & $-2iC$ & 0\tabularnewline
\hline 
\end{tabular}
\label{Comm1}
\vspace{5mm}
%
%
\caption{Commutation relations $[X,Y]$ in the $\left(L_{+},\vec{P}_{+},H_{\omega},\vec{P}_{-},L_{-}\right)$ basis.}
\begin{tabular}{|c|c|c|c|c|c|}
\hline 
\hspace{5mm}$X\backslash Y$\hspace{5mm} & \hspace{5mm}$P_{+j}$\hspace{5mm} & \hspace{5mm}$P_{-j}$\hspace{5mm} & \hspace{5mm}$\frac{H_{\omega}}{\omega}$\hspace{5mm} & \hspace{5mm}$L_{-}$\hspace{5mm} & \hspace{5mm}$L_{+}$\hspace{5mm} \tabularnewline
\hline 
$P_{+i}$ & 0 & $-\delta_{ij}N$ & $-P_{+i}$ & $-P_{-i}$ & 0\tabularnewline
\hline 
$P_{-i}$ & $\delta_{ij}N$ & 0 & $P_{-i}$ & 0 & $P_{+i}$\tabularnewline
\hline 
$\frac{H_{\omega}}{\omega}$ & $P_{+j}$ & $-P_{-j}$ & 0 & $-2L_{-}$ & $2L_{+}$\tabularnewline
\hline 
$L_{-}$ & $P_{-j}$ & 0 & $2L_{-}$ & 0 & $\frac{H_{\omega}}{\omega}$\tabularnewline
\hline 
$L_{+}$ & 0 & $-P_{+j}$ & $-2L_{+}$ & $-\frac{H_{\omega}}{\omega}$ & 0\tabularnewline
\hline 
\end{tabular}
\label{Comm2}
\vspace{5mm}
%
%
%
\caption{Commutation relations $[X,Y]$ in the $\left(Q_{+},\vec{P}_{+},H_{\omega},\vec{P}_{-},Q_{-}\right)$ basis.}
\begin{tabular}{|c|c|c|c|c|c|}
\hline 
\hspace{5mm}$X\backslash Y$\hspace{5mm} & \hspace{5mm}$P_{+j}$\hspace{5mm} & \hspace{5mm}$P_{-j}$\hspace{5mm} & \hspace{5mm}$\frac{H_{\omega}}{\omega}$\hspace{5mm} & $Q_{-}$ & $Q_{+}$\tabularnewline
\hline 
$P_{+i}$ & 0 & $-\delta_{ij}N$ & $-P_{+i}$ & 0 & 0\tabularnewline
\hline 
$P_{-i}$ & $\delta_{ij}N$ & 0 & $P_{-i}$ & 0 & 0\tabularnewline
\hline 
$\frac{H_{\omega}}{\omega}$ & $P_{+j}$ & $-P_{-j}$ & 0 & $-2Q_{-}$ & $2Q_{+}$\tabularnewline
\hline 
$Q_{-}$ & 0 & 0 & $2Q_{-}$ & 0 & $\frac{H_{\omega}}{\omega}$-$\frac{1}{2N}\left(\vec{P}_{+}\cdot\vec{P}_{-}+\vec{P}_{-}\cdot\vec{P}_{+}\right)$\tabularnewline
\hline 
$Q_{+}$ & 0 & $0$ & $-2Q_{+}$ & $-\frac{H_{\omega}}{\omega}+\frac{1}{2N}\left(\vec{P}_{+}\cdot\vec{P}_{-}+\vec{P}_{-}\cdot\vec{P}_{+}\right)$ & 0\tabularnewline
\hline 
\end{tabular}
\label{Comm3}
\end{table}

\section{Useful Taylor expansions}
\label{app:Taylor}

The following expansions are in either $x_{1}$ or $t_{1}$ about the origin. 
\begin{eqnarray}
\text{exp}\left[Ax_{1}\cdot x_{2}+\frac{B}{2}x_{1}^{2}\right]&=&\sum_{q=0}^{\infty}\sum_{n=0}^{\left\lfloor q/2\right\rfloor }\frac{1}{2^{n}n!(q-2n)!}A^{q-2n}B^{n}\left(x_{1}\cdot x_{2}\right)^{q-2n}\left(x_{1}^{2}\right)^{n}\label{Taylor1} \\
t_{21}^{-B}\text{exp}\left[\frac{C}{t_{21}}\right]&=&\text{exp}\left[\frac{C}{t_{2}}\right]\sum_{p=0}^{\infty}\frac{1}{p!}t_{1}^{p}t_{2}^{-B-2p}\sum_{r=0}^{p}\begin{pmatrix}p\\
r
\end{pmatrix}\frac{\Gamma(B+p)}{\Gamma(B+p-r)}C^{p-r}t_{2}^{r}\label{Taylor2} \\
t_{21}^{-B}\text{exp}\left[C\frac{t_{1}}{t_{21}}\right]&=&\sum_{p=0}^{\infty}\frac{1}{p!}t_{1}^{p}t_{2}^{-B-p}\sum_{r=0}^{p}\begin{pmatrix}p\\
r
\end{pmatrix}\frac{\Gamma(B+p)}{\Gamma(B+p-r)}C^{p-r}\label{Taylor3}
\end{eqnarray}

\bibliography{ReferencesBibtex}{}
\bibliographystyle{h-physrev5}

\end{document}